\documentclass[a4paper,12pt]{article}
\usepackage[dvips]{graphicx}
\usepackage{amsmath}        
\usepackage{amssymb}        
\usepackage{cite}

\begin{document}

\title{New Mixing Structures of Chiral Generations in a Model with
Noncompact Horizontal Symmetry} 

\author{Naoki Yamatsu\footnote{Electronic address: nyamatsu@indiana.edu}\\
{\it\small Department of Physics, Indiana University, Bloomington, IN 47405, U.S.A.}}
\date{ }

\maketitle

\begin{abstract}
 New mixing structures between chiral generations of elementary
 particles at low energy are shown in a vectorlike model with a
 horizontal symmetry $SU(1,1)$. In this framework the chiral model
 including odd number chiral generations is realized via the spontaneous
 symmetry breaking of the horizontal symmetry. 
 It is shown that the Yukawa coupling matrices of chiral generations
 have naturally hierarchical patterns, and in some cases the overall
 factors of their Yukawa coupling matrices, e.g. the Yukawa
 coupling constants of the bottom quark and tau lepton are naturally
 suppressed. 
\end{abstract}

\section{Introduction}

One of the most remarkable phenomena in Nature at low energies is the
existence of three chiral generations of quarks and leptons and their
hierarchical mass structures. 
There have been many attempts to understand the origin of
generations and/or the hierarchical structures by using various models;
e.g. models with horizontal symmetries $G_H$
\cite{Wilczek:1978xi,Froggatt:1978nt,Yanagida:1979as,Maehara:1979kf},  
which govern the generational structures of quarks and leptons;
grand unified models on the orbifold $S^1/Z_2$ 
\cite{Kawamura:2007cm,Kawamura:2009gr};
nonlinear $\sigma$ models on the coset space
$E_{7(+7){\rm global}}/SU(5)\times U(1)^{3}$  
\cite{Kugo:1983ai,Kugo:2010fs};
and magnetized orbifolding models 
\cite{Abe:2008fi,Abe:2008sx,Kobayashi:2010an}.

We focus on approaches based on horizontal symmetries. These symmetries
can be classified into three categories.
First, models with non-abelian group symmetries
\cite{Ishimori:2010au,King:2001uz} 
give an explanation for the three generations of quarks and leptons if
the group involves a triplet representation.
Second, models with an abelian group $U(1)$
\cite{Froggatt:1978nt,Maekawa2004} 
naturally gives an account for the hierarchical mass structures 
of the three chiral generations of quarks and leptons by using the
Froggatt-Nielsen mechanism \cite{Froggatt:1978nt}. 
Third, a model with a noncompact nonabelian group symmetry $SU(1,1)$
\cite{Inoue:1994qz}
allows us the opportunity that chiral generations of
elementary particles and their hierarchical mass structures are
understood by using the spontaneous symmetry breaking, one of the
essential concepts of modern particle physics, of the horizontal symmetry,
where the noncompact group $SU(1,1)$ is a special pseudo-unitary group
\cite{Gourdin:1967,Gilmore:102082}. 
(Another example of a special pseudo-unitary group is  the Lorentz group
$SO(3,1)$.)

The author has previously discussed an $\mathcal{N}=1$ supersymmetric
vectorlike model 
with a noncompact group $SU(1,1)$ horizontal symmetry
\cite{Yamatsu:2007,Yamatsu:2008}.
This model has some important features:
chiral gauge theories derived from vectorlike gauge theories
\cite{Inoue:1994qz};
the hierarchical structure of Yukawa coupling constants of chiral
matter at low energy \cite{Yamatsu:2007};
and the spontaneous breakdown of P, C and T symmetries
\cite{Yamatsu:2008}.
Once we apply this model to the standard model (SM),  these features
can realize almost the constrained minimal supersymmetric standard
model (MSSM) \cite{Martin:1997ns,Baer2006} at low energy
\cite{Yamatsu:2008}. 
See Ref.~\cite{Yamatsu:2011} for a review. 

The main purpose of this paper is to show new types of structures to
realize 
chiral generations of elementary particles at low energy in a model with
the horizontal symmetry $SU(1,1)$. 
In some cases the pattern of the Yukawa coupling constants of the chiral
generations derived from the new structures is much different
from that discussed before. 
The difference between the Yukawa coupling constants of top quarks,
bottom quarks and tau leptons is also discussed.
In \S \ref{sec:Overview} we give an overview of the noncompact
horizontal symmetry and define the terms of the model.
In \S \ref{sec:SGG} we discuss how to realize chiral generations in the
context of  the known structure and the new structure.
In \S \ref{sec:Yukawa} we analyze typical Yukawa coupling structures 
by using the mixing patterns of chiral generations in \S \ref{sec:SGG}.
Section \ref{summary} is devoted to summary and discussion.

\section{Overview of the Model}
\label{sec:Overview}

We give an overview of the noncompact horizontal
symmetry $SU(1,1)$ and define the terms of the model. 
Let us begin by defining the three generators $\tau_a$
$(a=1,2,3)$ of $SU(1,1)$ satisfying the commutation relations 
\begin{eqnarray}
[\tau_1,\tau_2]=-i\tau_3,\ \ \ 
[\tau_2,\tau_3]=+i\tau_1,\ \ \ 
[\tau_3,\tau_1]=+i\tau_2.
\end{eqnarray}
The components of any representation of $SU(1,1)$ are labeled by
the  Casimir operator $\mathbf{\tau}^2$ of $SU(1,1)$ and the weight of the
third component $\tau_3$ of $SU(1,1)$, where the Casimir operator of
$SU(1,1)$ is defined by 
$\mathbf{\tau}^2:= \tau_1^2+\tau_2^2-\tau_3^2$.

We use two types of representations of $SU(1,1)$; one type is 
unitary infinite-dimensional representations, which 
are constructed using all Hermitian generators $\tau_a$;
the other type is nonunitary finite-dimensional representations,
which are constructed using two anti-Hermitian generators
$\tau_1$ and $\tau_2$ and one Hermitian generator $\tau_3$.
Two types of infinite-dimensional representations are used; one
representation has only positive weights of the third 
component generator $\tau_3$, where the lowest state is vanished by the
ladder operator $\tau_{-}:=\tau_1-i\tau_2$; 
the other has only negative weights of the third 
component generator $\tau_3$, where the highest state is
vanished by the ladder operator $\tau_{+}:=\tau_1+i\tau_2$.
A field in the representation with only a positive weight 
is referred to as a positive field denoted by, e.g., 
$\hat{F}=\{\hat{f}_\alpha,\hat{f}_{\alpha+1},\cdots\}$,
where the subscripts of the components $\hat{f}_{\alpha+i}$
$(i=0,1,2,\cdots)$ of $\hat{F}$ stand for the weight of $\tau_3$ of
$SU(1,1)$. 
A field in the representation with only a negative weight is
referred as a negative field denoted by, for example, 
$\hat{F}^c=\{\hat{f}^c_{-\alpha},\hat{f}^c_{-\alpha-1},\cdots\}$.
The finite-dimensional representations are characterized by the highest
weight $S$ referred to as the $SU(1,1)$ spin $S$ and the weight of
eigenvalue of the third component generator $\tau_3$, where 
the highest state is vanished by the operator $\tau_{+}$, and the lowest
state is vanished by the operator $\tau_{-}$. 
A field in a finite-dimensional representation is referred to as a
finite field denoted by, e.g.,
$\hat{\Psi}=\{\hat{\psi}_{-S},\hat{\psi}_{-S+1},\cdots,
\hat{\psi}_{S-1},\hat{\psi}_S\}$,
where the subscripts of the components $\hat{\psi}_{n}$
$(n=-S,-S+1,\cdots,S)$ of the finite field $\hat{\Psi}$ stand for the
weight of $\tau_3$, and $S$ is a non-negative integer or a half-integer. 
In the following, we refer to the positive and negative fields
introduced in a vectorlike manner as matter fields, such as quarks,
leptons and higgses. The finite fields are referred to as structure
fields because these determine the generational structures of the model
and do not correspond to the SM or supersymmetric SM fields. 

Before we finish this section, we introduce two cubic invariants under
$SU(1,1)$ transformations \cite{Yamatsu:2007}; one is built from two
infinite dimensional unitary representations with one positive and one
negative weight and finite-dimensional representations; the other
consists of three infinite-dimensional unitary representations with two
positive weights and one negative weight, or one positive weight and two
negative weights. The former is the following cubic coupling term
\begin{equation}
\hat{F}\hat{G}^c\hat{\Psi}=\sum_{i,j=0}^\infty
D^{\beta,\alpha,S}_{j,i}\ \hat{f}_{\alpha+i}\hat{g}^c_{-\beta-j}
\hat{\psi}_{-i+j-q},
\end{equation}
where $\alpha(-\beta)$ is the lowest(highest) weight of the matter field
$\hat{F}(\hat{G}^c)$, and $q$ is defined as $q:=\alpha-\beta$
and is an integer or half-integer, which is allowed when the $SU(1,1)$
spin $S$ of the structure field $\hat{\Psi}$ satisfies $S\geq |q|$. 
For $S\geq |-i+j-q|$, the Clebsch-Gordan coefficient (CGC) is 
\begin{align}
D^{\beta,\alpha,S}_{j,i}&=(-1)^i
\sqrt{\frac{i!j!(i-j+S+q)!(-i+j+S-q)!}
{\Gamma(2\alpha+i)\Gamma(2\beta+j)}}\nonumber\\
&\hspace{-3em}\times\sum_{r=0}^{S+q}\frac{\Gamma(2\beta+j+r)}
{(S+q-r)!(i-j+S+q-r)!r!
(-i+j-2q+r)!(j-S-q+r)!}~;
\label{CGC-D}
\end{align}
otherwise, $D^{\beta,\alpha,S}_{j,i}=0$.
The CGC $D_{j,i}^{\beta,\alpha,S}$ satisfies the symmetric relation
$D_{j,i}^{\beta,\alpha,S}=(-1)^{i-j}D_{i,j}^{\alpha,\beta,S}$.
The CGC $D_{j,i}^{\beta,\alpha,S}$ of the matter and structure fields
$\hat{F}\hat{F}^c\hat{\Psi}$ behaves as $i^S$ in the limit 
$i,j\to \infty$ with $|i-j|$ fixed:
\begin{align}
D_{j,i}^{\beta,\alpha,S}\simeq
(-1)^ii^S\frac{(2S)!}
{(S+q)!(S-q)!\sqrt{(i-j+S-q)!(-i+j+S+q)!}}.
\label{D_asymptotic}
\end{align}
As we find in \S \ref{sec:SGG}, this asymptotic behavior is essential
to realize chiral theories from vectorlike theories.
The latter is 
\begin{equation}
\hat{F}\hat{G}\hat{H}=
\sum_{i,j=0}^\infty
C_{i,j}^{\alpha,\beta,\Delta}
\hat{f}_{\alpha+i}\hat{g}_{\beta+j}\hat{h}_{-\gamma-i-j+\Delta},
\end{equation}
where  $\Delta$ is a semi-positive integer $(\Delta\geq 0)$.
For $i+j<\Delta$, the CGC is zero and, for $i+j\geq\Delta$,
\begin{align}
C_{i,j}^{\alpha,\beta,\Delta}=&
(-1)^{i+j}\sqrt{\frac{i!j!}{(i+j-\Delta)!}
\frac{\Gamma(2\alpha+i)\Gamma(2\beta+j)}
{\Gamma(2\gamma+i+j+\Delta)}}\nonumber\\
&\hspace{-3em}
\times \sum_{r=0}^\Delta(-1)^r
\frac{(i+j-\Delta)!\Gamma(2\alpha)\Gamma(2\beta)}
{(i-r)!(j+r-\Delta)!r!(\Delta-r)!\Gamma(2\alpha+r)
\Gamma(2\beta+\Delta-r)},
\label{CGC-C}
\end{align}
where the $C_{i,j}^{\alpha,\beta,\Delta}$ satisfies the symmetric
relation 
$C_{i,j}^{\alpha,\beta,\Delta}=(-1)^\Delta C_{j,i}^{\beta,\alpha,\Delta}$.

\section{Spontaneous Generation of Generations}
\label{sec:SGG}

We now discuss how to extract chiral matter content from vectorlike matter
content. This mechanism is referred to as the spontaneous generation of 
generations \cite{Inoue:1994qz}.
The mechanism can produce finite numbers of chiral generations of the
SM fields at low energies, such as quarks, from matter fields, where
matter fields belong to the infinite-dimensional representation of
$SU(1,1)$ in a vectorlike manner. The appearance of chiral generations
of matter is dominantly dependent on the $SU(1,1)$ spins of the
structure fields $\hat{\Psi}$s with non-vanishing vacuum expectation
values (VEVs) and subdominantly depends on certain combinations of the 
VEVs of the structure fields $\hat{\Psi}$s and coupling constants,
because of normalizable condition of chiral particles.

We first investigate the superpotential that includes 
two structure fields with integer spins discussed in
Ref.~\cite{Inoue:2000ia,Yamatsu:2007}. 
After that, the superpotential that includes a structure field with a
half-integer spin is discussed.

\subsection{Two Structure Fields with an Integer Spin}

We consider the superpotential that contains 
the matter fields $\hat{F}$ and $\hat{F}^c$ and the structure fields
$\hat{\Psi}$ and $\hat{\Psi}^{\prime}$ 
\begin{align}
W=x\hat{F}\hat{F}^c\hat{\Psi}
+x'\hat{F}\hat{F}^c\hat{\Psi}^{\prime},
\label{W-FFbar}
\end{align}
where $x$ and $x'$ are real coupling constants.
We assume non-vanishing VEVs for the $0$th and $-g$th components of
the structure fields $\hat{\Psi}$ and $\hat{\Psi}^{\prime}$
with the $SU(1,1)$ spins $S$ and $S'$, 
respectively, and assume that the scale of these VEVs is a high
energy scale such as the GUT scale or Planck scale and the typical
energy  scale of massless matter is a low scale such as the
electroweak scale. The non-vanishing VEVs $\langle\psi_{0}\rangle$ and 
$\langle\psi_{-g}'\rangle$ can generate $g$ massless modes 
$\{\hat{f}_0, \hat{f}_1, \hat{f}_2, \cdots, \hat{f}_{g-1} \}$ 
$(S'\geq g)$ from the positive field $\hat{F}$ with the form 
\begin{align}
\hat{f}_{\alpha+i}=
\sum_{n=0}^{g-1} \hat{f}_n U_{n,i}^f + [\mbox{massive modes}],
\label{f-massless-mode}
\end{align}
where $\hat{f}_{\alpha+i}$ $(i=0,1,2,\cdots)$ are the components of the
positive field $\hat{F}$ whose lowest weight is $\alpha$,
$\hat{f}^c_{-\alpha-i}$ are only massive modes, and $U_{n,i}^f$
is referred to as the mixing coefficient of the matter field
$\hat{F}$. The coefficient represents the relation between the
eigenstates of the third component generator of $SU(1,1)$ and mass
eigenstates of massless modes.  

Let us calculate the mass term of the superpotential in
Eq.~(\ref{W-FFbar}) at the vacuum to obtain the massless and chiral
matter fields. 
Substituting the modes in Eq.~(\ref{f-massless-mode}) for the
superpotential in Eq.~(\ref{W-FFbar}), we obtain 
\begin{align}
\left.W\right|_{\Psi=\langle\Psi\rangle}
&=x\hat{F}\hat{F}^c\langle\hat{\Psi}\rangle
+x'\hat{F}\hat{F}^c\langle\hat{\Psi}^{\prime}\rangle\nonumber\\
&=\sum_{i=0}^\infty
\left(xD_{i,i}^{\alpha,\alpha,S}\langle\psi_{0}\rangle
 \hat{f}_{\alpha+i}\hat{f}^c_{-\alpha-i} 
+x'D_{i,i+g}^{\alpha,\alpha,S'}\langle\psi_{-g}^{\prime}\rangle
\hat{f}_{\alpha+i+g}\hat{f}^c_{-\alpha-i}\right)\nonumber\\
&=\sum_{i=0}^\infty \hat{f}_{n}
\left(xD_{i,i}^{\alpha,\alpha,S}\langle\psi_{0}\rangle U_{n,i}^f
+x'D_{i,i+g}^{\alpha,\alpha,S'}\langle\psi_{-g}^{\prime}\rangle
U_{n,i+g}^f\right)\hat{f}^c_{-\alpha-i}
+[\mbox{massive modes}], 
\label{F-mass}
\end{align}
where $D_{i,i}^{\alpha,\alpha,S}$ and $D_{i,i+g}^{\alpha,\alpha,S'}$
are CGCs of the positive-negative-finite
field coupling defined in Eq.~(\ref{CGC-D}).

The massless modes $\hat{f}_n$ are extracted from the component
$\hat{f}_{\alpha+i}$ of the matter field $\hat{F}$.
The orthogonality of the massless modes $\hat{f}_n$ to the massive modes 
$\hat{f}_{-\alpha-i}^c$
requires the coefficients $U_{n,i}^f$ to satisfy the recursion equation   
\begin{equation}
x\langle\psi_0\rangle D_{i,i}^{\alpha,\alpha,S}U_{n,i}^f
+x'\langle\psi_{-g}^{\prime}\rangle
D_{i,i+g}^{\alpha,\alpha,S'}U_{n,i+g}^f=0.
\label{recursion_equation_minimal}
\end{equation}
This equation gives the mixing coefficients
\begin{equation}
U^f_{n,i}=U^f_n\sum_{s=0}^\infty\delta_{i,n+gs}
(-\epsilon)^s b^f_{n,s},\ \ \ \ 
\epsilon=
\frac{x\langle\psi_{0}\rangle}{x'\langle\psi_{-g}^{\prime}\rangle},\ \ \
b^f_{n,s}=\prod_{r=0}^{s-1}
\frac{D^{\alpha,\alpha,S}_{gr+n,gr+n}}
{D^{\alpha,\alpha,S'}_{gr+n,g(r+1)+n}},
\label{mixing-form}
\end{equation}
where $n=0,1,2,\cdots g-1$, and the parameter $\epsilon$ depends on the
couplings and the VEVs. The $b_{n,s}^f$ is determined by the weights of
the $SU(1,1)$ $\alpha$ and $S$, where $b_{n,0}^f:=1$.

\begin{figure}
\centering
\includegraphics[totalheight=3cm]{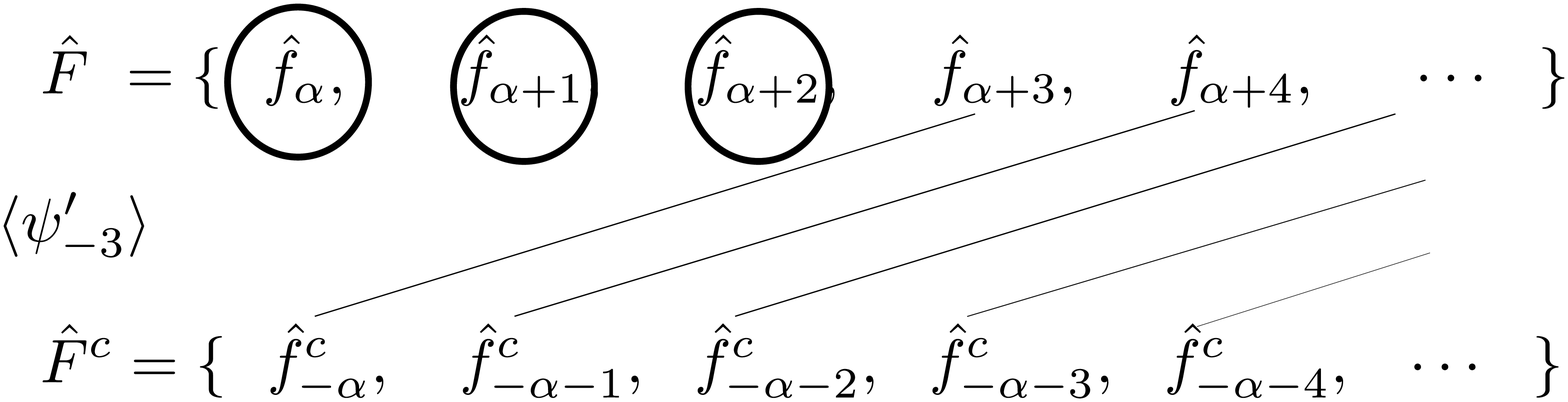}
\caption{An example for the three massless generation case assuming
the VEVs $\langle\psi_0\rangle=0$ and $\langle\psi_{-g}'\rangle\not=0$
and $g=3$ in Eq.~(\ref{F-mass}):
the components $\hat{f}_{\alpha}$, $\hat{f}_{\alpha+1}$ and
 $\hat{f}_{\alpha}$ of the matter field $\hat{F}$ are massless, and the
 other components of $\hat{F}$ and all the components of $\hat{F}^c$ are
 massive,
where the massless modes are surround by a circle, and 
the components $\hat{f}_{\alpha+3+i}$ and $\hat{f}_{-\alpha-i}^c$
$(i=0,1,2,\cdots)$ of the matter fields $\hat{F}$ and $\hat{F}^c$
connected by a solid line have a mass term
that comes from the VEV $\langle\psi_{-3}'\rangle$.}
\label{fig:SGG-min-g=3}
\end{figure}

\begin{figure}
\centering
\includegraphics[totalheight=3cm]{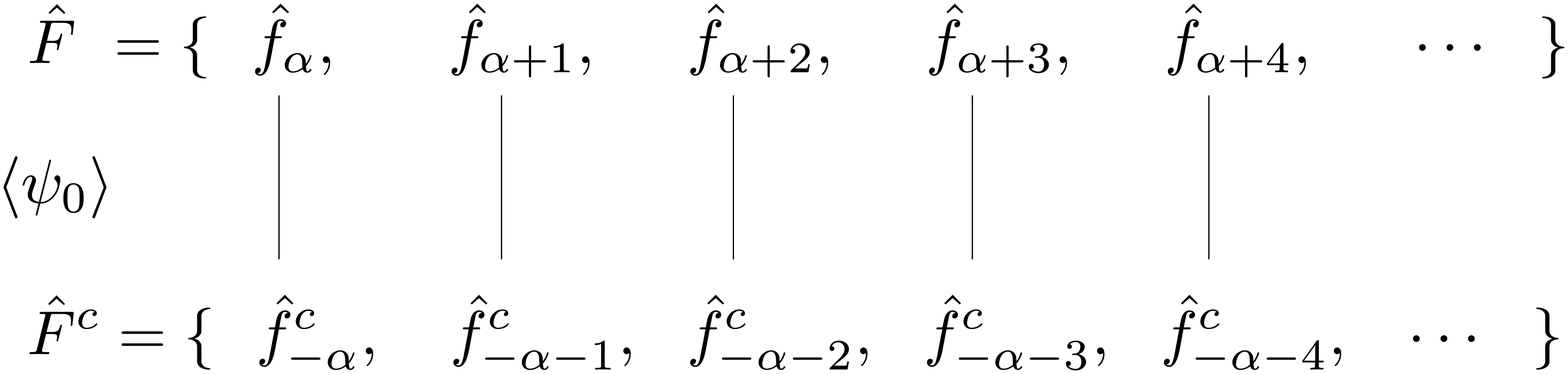}
\caption{An example for the no massless generation case assuming
the VEVs $\langle\psi_0\rangle\not=0$ and $\langle\psi_{-g}'\rangle=0$
in Eq.~(\ref{F-mass}):
all the components of the matter fields $\hat{F}$ and $\hat{F}^c$ are
massive, 
where all the components $\hat{f}_{\alpha+i}$ and
$\hat{f}_{-\alpha-i}^c$ $(i=0,1,2,\cdots)$ of the matter fields
$\hat{F}$ and $\hat{F}^c$ connected by a solid
line have a mass term that comes from the VEV $\langle\psi_{0}\rangle$.}
\label{fig:SGG-min-g=0}
\end{figure}

To understand the meaning of the mixing coefficient in
Eq.~(\ref{mixing-form}), we first examine two extreme cases of VEVs;
$\langle\psi_0\rangle=0$ and $\langle\psi_{-g}'\rangle\not=0$;
and $\langle\psi_0\rangle\not=0$ and $\langle\psi_{-g}'\rangle=0$.
In the case of the VEVs $\langle\psi_0\rangle=0$ and
$\langle\psi_{-g}'\rangle\not=0$, 
the recursion equation in Eq.~(\ref{recursion_equation_minimal})
constrains the mixing coefficients $U_{n,i+g}^f=0$
$(i=0,1,2,\cdots)$ and does not determine $U_{n,k}^f$
$(k=1,2,\cdots,g-1)$, and we take a normalization condition
$U_{n,i}^f=\delta_{n,i}$. 
This means that the component $\hat{f}_{\alpha+n}$ $(n=0,1,\cdots,g-1)$
can be identified as its corresponding massless mode $\hat{f}_n$.
For example, for $g=3$, from Fig.~\ref{fig:SGG-min-g=3}, 
since the components $\hat{f}_{\alpha}$, $\hat{f}_{\alpha+1}$ and
$\hat{f}_{\alpha+2}$ do not have any mass term, they are massless.
The other components $\hat{f}_{\alpha+3+i}$ and
$\hat{f}_{-\alpha-i}^c$ $(i=0,1,2,\cdots)$ have a corresponding mass
$x'D_{i,i+3}^{\alpha,\alpha,S'}\langle\psi_{-3}'\rangle$, 
they are massive. Three massless modes appear, and 
they can be identified as $\hat{f}_0=\hat{f}_\alpha$,
$\hat{f}_1=\hat{f}_{\alpha+1}$ and $\hat{f}_2=\hat{f}_{\alpha+2}$.
Next, in the case of the VEVs $\langle\psi_0\rangle\not=0$ and
$\langle\psi_{-g}'\rangle=0$,
the recursion equation in Eq.~(\ref{recursion_equation_minimal})
determines the mixing coefficients satisfying $U_{n,i}^f=0$
$(i=0,1,2,\cdots)$ and from Fig.~\ref{fig:SGG-min-g=0}, 
since all the components $\hat{f}_{\alpha+i}$ and
$\hat{f}_{-\alpha-i}^c$ have a mass term, no massless modes emerge.

\begin{figure}
\centering
\includegraphics[totalheight=3cm]{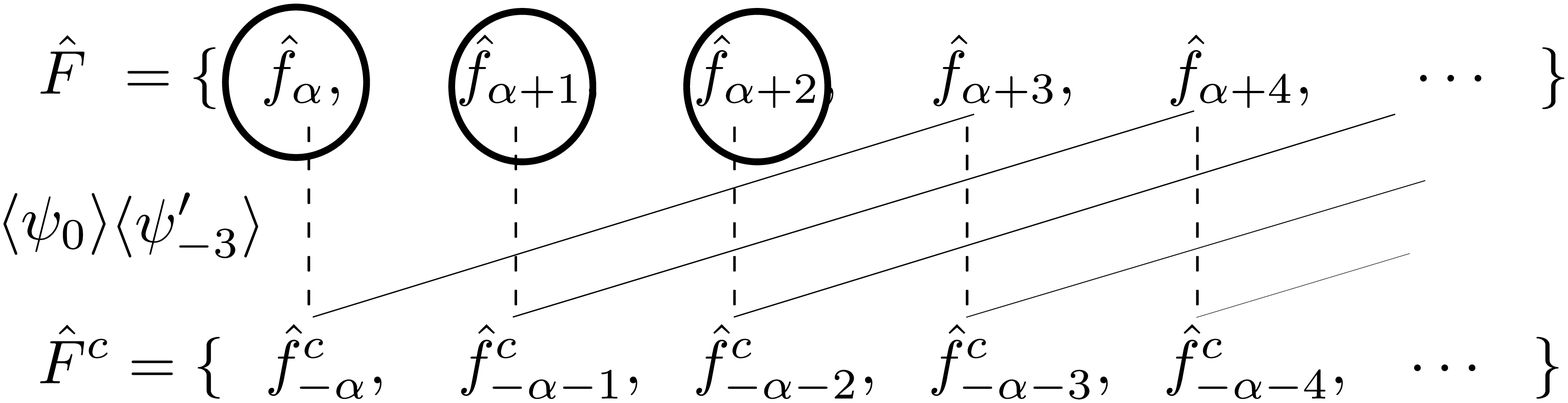}
\caption{An example for the three massless generation case $g=3$ in
Eq.~(\ref{F-mass}):
each massless mode $\hat{f}_n$ $(n=0,1,2)$ is realized by
certain linear combinations of the components $\hat{f}_{\alpha+3k+n}$
$(k=1,2,\cdots)$, and the constructional element of each massless mode
is determined by its mixing coefficients $U_{n,n+3k}^f$ given in
Eq.~(\ref{mixing-form}), 
where the components of the matter fields $\hat{F}$ and $\hat{F}^c$
connected by solid and dashed lines have mass terms that 
come from the VEVs $\langle\psi_{-3}'\rangle$ and
$\langle\psi_{0}\rangle$, respectively,
and the mass term of the solid line that comes from the VEV
$\langle\psi_{-3}'\rangle$ dominantly contributes to whether
massless modes appear or not.}
\label{fig:SGG-min-g=3-v2}
\end{figure}

\begin{figure}
\centering
\includegraphics[totalheight=3cm]{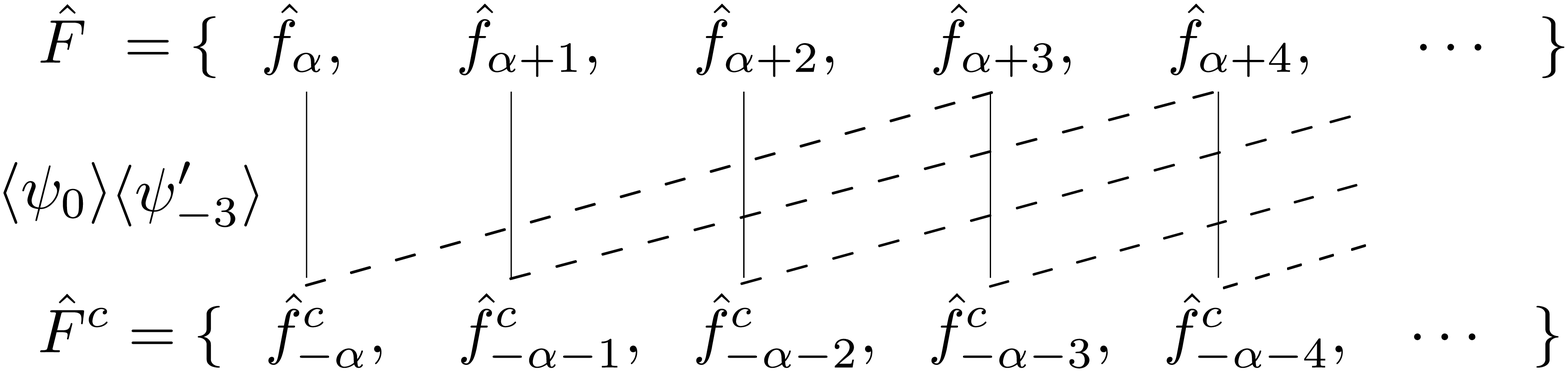}
\caption{An example for the no massless generation case $g=3$ in
Eq.~(\ref{F-mass}),
where the components of the matter fields $\hat{F}$ and $\hat{F}^c$
connected by solid and dashed lines have mass terms that  
comes from the VEVs $\langle\psi_{0}\rangle$ and
$\langle\psi_{-3}'\rangle$, respectively,
and the mass term of the solid line that comes from the VEV
$\langle\psi_{0}\rangle$ dominantly contributes to whether
massless modes appear or not.}
\label{fig:SGG-min-g=0-v2}
\end{figure}

Let us move on to the case of the VEVs $\langle\psi_0\rangle\not=0$
and $\langle\psi_{-g}'\rangle\not=0$ in
Eqs.~(\ref{recursion_equation_minimal}) and (\ref{mixing-form}).
From the above consideration for the cases of $\langle\psi_0\rangle=0$ or
$\langle\psi_{-g}'\rangle=0$, 
in the case of the VEVs $\langle\psi_0\rangle\not=0$
and $\langle\psi_{-g}'\rangle\not=0$, we expect that the massless mode
$\hat{f}_n$ $(n=0,1,\cdots,g-1)$ that is realized by certain linear
combinations of the components $\hat{f}_{\alpha+gk+n}$ $(k=1,2,\cdots)$
appears as in Fig.~\ref{fig:SGG-min-g=3-v2} or no massless mode
appears as in Fig.~\ref{fig:SGG-min-g=0-v2}. 
In fact the realization of the massless modes $\hat{f}_n$
$(n=0,1,\cdots,g-1)$ requires a normalizable condition
$\sum_{i=0}^\infty |U_{n,i}^f|^2<\infty$;
if this condition is not satisfied, the massless modes $\hat{f}_n$ are
illusions without any physical reality. 
Since the mixing coefficients in Eq.~(\ref{mixing-form}) are a geometric
series, the requirement of the normalizable condition corresponds to the
condition 
\begin{equation}
\lim_{i\rightarrow\infty}\left|\frac{U_{n,i+g}^f}{U_{n,i}^f}\right|<1.
\label{normalizable_condition}
\end{equation}
By using the asymptotic behavior of the CGC $D_{j,i}^{\beta,\alpha,S}$
in Eq.~(\ref{D_asymptotic}), we find 
\begin{equation}
\lim_{i\rightarrow\infty}
\left[\left|\frac{U_{n,i+g}^f}{U_{n,i}^f}\right|
-\left|\frac{\epsilon}{\epsilon_{cr}}\right| 
i^{S-S'}\right]=0,
\end{equation}
where $\epsilon_{cr}$ is a constant that is dependent on the $SU(1,1)$
spins and the number of generations and is independent of the component
$i$. Thus, the requirement of the normalizable condition is dominantly
determined by the relation between the $SU(1,1)$ spins $S$ and $S'$ and
subdominantly depends on the relation between $\epsilon$ and
$\epsilon_{cr}$. 

Whether massless generations appear or not  can be classified into three
types by using the $SU(1,1)$ spins $S$ and $S'$
of the structure fields $\hat{\Psi}$ and $\hat{\Psi}'$, respectively:
$S>S'$, $S<S'$ and $S=S$. First, for $S>S'$, since
$\lim_{i\to\infty}|U_{n,i+g}^f/U_{n,i}^f|=0$,
the normalizable condition in Eq.~(\ref{normalizable_condition}) is
always satisfied. 
Thus, the massless modes $\hat{f}_n$ appear as in
Fig.~\ref{fig:SGG-min-g=3-v2} for any value of the parameter $\epsilon$.  
Next, for $S<S'$, since
$\lim_{i\to\infty}|U_{n,i+g}^f/U_{n,i}^f|=\infty$, 
the normalizable condition in Eq.~(\ref{normalizable_condition}) is not
satisfied. 
Thus, no massless mode appear as in
Fig.~\ref{fig:SGG-min-g=0-v2} for any value of the parameter $\epsilon$.
Finally, for $S=S'$, whether the normalizable condition is satisfied 
and which situation occurs as in Fig.~\ref{fig:SGG-min-g=3-v2} or
Fig.~\ref{fig:SGG-min-g=0-v2} depend on the parameter $\epsilon$ because 
$\lim_{i\to\infty}|U_{n,i+g}^f/U_{n,i}^f|=|\epsilon/\epsilon_{cr}|$. 
The condition $S=S'$ is referred to as a marginal assignment. 
The critical value $\epsilon_{cr}$ can be calculated by using the mixing
coefficient in Eq.~(\ref{mixing-form}) and the asymptotic form of the
CGC $D_{j,i}^{\beta,\alpha,S}$ in Eq.~(\ref{D_asymptotic});
the parameter $\epsilon$ must satisfy the constraint 
\begin{align}
|\epsilon| < \epsilon_{cr}=
\sqrt{\frac{S!S!}{(S+g)!(S-g)!}}
\label{epsilon-critical}
\end{align}
to produce the massless modes $\hat{f}_n$.

The chiral nature of supersymmetry plays an important role in this
model. If the model is not based on supersymmetry and matter fields
$F$ and $F^c$ are spin-$1/2$ fermions and a structure
field $\Psi'$ is a spin-$0$ boson, then a Yukawa coupling term 
$FF^c\Psi'$ is allowed by gauge symmetry and the conjugate term
$FF^c\Psi^{\prime \dag}$ is also permitted because the structure field
$\Psi'$ belongs to the real representation of $SU(1,1)$. The latter term  
destroys the chiral structures and all massless modes disappear at
low energy. Supersymmetry naturally forbids the latter coupling by
its chiral nature.

\subsection{One Structure Field with an Integer Spin and 
One with a Half-Integer Spin}

Let us start to discuss the structures of the massless modes in a model
with structure fields with an integer spin and a half-integer spin.
We introduce an additional set of the matter fields $\hat{F}'$ and
$\hat{F}'{}^c$ with the lowest weight $\alpha'$ and the highest weight
$-\alpha'$, respectively.
We choose the value 
\begin{equation}
q:=\alpha'-\alpha 
\end{equation}
to be a positive half-integer.
We suppose that massless generations are realized as a linear 
combination of the components of the matter fields  $\hat{F}$ and
$\hat{F}'$
\begin{equation}
\hat{f}_{\alpha+i}=\sum_{n=0}\hat{f}_nU^f_{n,i}+[{\rm massive\ modes}],
\ \ \ 
\hat{f}_{\alpha'+i}^{\prime}=\sum_{n=0}\hat{f}_nU_{n,i}^{f\prime}
+[{\rm massive\ modes}],
\label{eq:minimal_ext_massless_mode}
\end{equation}
and the conjugate fields $\hat{f}_{-\alpha-i}^c$ and
$\hat{f}_{-\alpha'-i}^{\prime c}$ do not include these massless modes,
where  $n$ is a label of these massless modes.
We also introduce two structure fields $\hat{\Phi}$ and $\hat{\Psi}$
with an $SU(1,1)$ integer spin $S$ and an $SU(1,1)$ half-integer spin 
$S'$, respectively. They couple to the matter fields in the following
form 
\begin{align}
W&=x\hat{F}\hat{F}^c\hat{\Phi}+x'\hat{F}'\hat{F}'{}^c\hat{\Phi}
+z\hat{F}'\hat{F}^{c}\hat{\Psi}+z'\hat{F}\hat{F}'{}^{c}\hat{\Psi},
\label{Wf-min}
\end{align}
where $x,x',z,z'$ are coupling constants and $q$ must be less than or
equal to $S'$ $(q\leq S')$ to realize the superpotential. 

We discuss how many generations of matter fields $\hat{f}_n$ are
produced from a double matter sector through the superpotential
in Eq.~(\ref{Wf-min}). 
We assume that the $0$th and $-p$th $(p>0)$ components of the structure
fields $\hat{\Phi}$ and $\hat{\Psi}$ acquire non-vanishing VEVs:
\begin{equation}
\langle\hat{\Phi}\rangle=\langle\phi_{0}\rangle,
\hspace{1em}
\langle\hat{\Psi}\rangle=\langle \psi_{-p}\rangle,
\label{VEV-min-ext}
\end{equation}
where $p$ must be less than or equal to $S'$ $(p\leq S')$.

We calculate the mass term of the superpotential
in Eq.~(\ref{Wf-min}) to confirm how much massless matter emerges in
the low energy physics. By substituting the expressions in
Eq.~(\ref{VEV-min-ext}) into the mass term in Eq.~(\ref{Wf-min}) at
the vacuum, we have 
\begin{align}
\left.W\right|_{\Psi=\langle\Psi\rangle}
&=x\hat{F}\hat{F}^c\langle\hat{\Phi}\rangle
+x'\hat{F}'\hat{F}'{}^c\langle\hat{\Phi}\rangle
+z\hat{F}'\hat{F}^{c}\langle\hat{\Psi}\rangle
+z'\hat{F}\hat{F}'{}^{c}\langle\hat{\Psi}\rangle
\nonumber\\
&=\sum_{i=0}^\infty
\bigg[xD_{i,i}^{\alpha,\alpha,S}\langle\phi_{0}\rangle
\hat{f}_{\alpha+i}\hat{f}_{-\alpha-i}^{c}
+x'D_{i,i}^{\alpha',\alpha',S}\langle\phi_{0}\rangle
\hat{f}_{\alpha'+i}^{\prime}\hat{f}_{-\alpha'-i}^{\prime c}\nonumber\\
&\hspace{2em} 
+zD_{i,i+p-q}^{\alpha,\alpha',S'}\langle\psi_{-p}\rangle
\hat{f}_{\alpha'+i+p-q}^{\prime}\hat{f}_{-\alpha-i}^{c}
+z'D_{i,i+p+q}^{\alpha',\alpha,S'}\langle\psi_{-p}\rangle
\hat{f}_{\alpha+i+p+q}\hat{f}_{-\alpha'-i}^{\prime c}\bigg]
\nonumber\\
&=\sum_{n=0}\sum_{i=0}^\infty 
\bigg[
\hat{f}_n\left(
xD_{i,i}^{\alpha,\alpha,S}\langle\phi_{0}\rangle U^f_{n,i}
+zD_{i,i+p-q}^{\alpha,\alpha',S'}
\langle\psi_{-p}\rangle U_{n,i+p-q}^{f\prime}
\right)\hat{f}_{-\alpha-i}^{c}\nonumber\\
&\hspace{0em}+\hat{f}_n\left(
x'D_{i,i}^{\alpha',\alpha',S}
\langle\phi_{0}\rangle U_{n,i}^{f\prime}
+z'D_{i,i+p+q}^{\alpha',\alpha,S'}
\langle\psi_{-p}\rangle U^f_{n,i+p+q}
\right)\hat{f}_{-\alpha'-i}^{\prime c}
+[\mbox{massive modes}]
\bigg].
\label{FF'-mass}
\end{align}
The emergence of the massless modes $\hat{f}_n$ requires that 
the coefficients of the terms of the massless modes $\hat{f}_n$ coupling
to the massive modes $\hat{f}_{-\alpha-n}^c$ and
$\hat{f}_{-\alpha-n}'^c$ must vanish simultaneously:
\begin{align}
&xD_{i,i}^{\alpha,\alpha,S}\langle\phi_{0}\rangle U_{n,i}^f
+zD_{i,i+p-q}^{\alpha,\alpha',S'}
\langle\psi_{-p}\rangle U_{n,i+p-q}^{f\prime}=0,
\label{FF'-mass-rec-1}\\
&x'D_{i,i}^{\alpha',\alpha',S}
\langle\phi_{0}\rangle U_{n,i}^{f\prime}
+z'D_{i,i+p+q}^{\alpha',\alpha,S'}
\langle\psi_{-p}\rangle U_{n,i+p+q}^f=0.
\label{FF'-mass-rec-2}
\end{align}
These recursion equations lead to the mixing coefficients
\begin{align}
U_{n,i}^f&=U_{n,\ell}^f\sum_{s=0}^\infty 
\delta_{i,\ell+2ps}\epsilon^s b_{\ell,s}^f,\ 
b_{\ell,s}^f=\prod_{r=0}^{s-1}
\frac{D_{\ell+2pr,\ell+2pr}^{\alpha,\alpha,S}
D_{\ell+2pr+p-q,\ell+2pr+p-q}^{\alpha',\alpha',S}}
{D_{\ell+2pr+p-q,\ell+2p(r+1)}^{\alpha',\alpha,S'}
D_{\ell+2pr,\ell+2pr+p-q}^{\alpha,\alpha',S'}},\ 
\epsilon=
\frac{xx'\langle\phi_{0}\rangle^2}{zz'\langle\psi_{-p}\rangle^2},
\label{mixing-coefficient-half-1}\\
U_{n,j}^{f\prime}&=U_{n,\ell'}^{f\prime}\sum_{s=0}^\infty 
\delta_{j,\ell'+2ps}\epsilon^s b_{\ell',s}^{f\prime},\
b_{\ell',s}^{f\prime}=\prod_{r=0}^{s-1}
\frac{D_{\ell'+2pr,\ell'+2pr}^{\alpha',\alpha',S}
D_{\ell'+2pr+p+q,\ell'+2pr+p+q}^{\alpha,\alpha,S}}
{D_{\ell'+2pr,\ell'+2pr+p+q}^{\alpha',\alpha,S'}
D_{\ell'+2pr+p+q,\ell'+2p(r+1)}^{\alpha,\alpha',S'}},
\label{mixing-coefficient-half-2}
\end{align}
where $\ell$ and $\ell'$ are defined as the lowest states of the matter
fields $\hat{F}$ and $\hat{F}'$ of each massless mode, and
$b_{\ell^{(\prime)},0}^{f(\prime)}=1$. 

We must determine the relation between the initial conditions
$U_{n,\ell}^f$ and $U_{n,\ell'}^{f\prime}$.
The relation  can be classified into three conditions:
Type-I, $p>q$; Type-II, $p=q$; and Type-III, $p<q$. 
For the Type-I condition $p>q$, $n$ is defined as 
$n=\ell=\ell'-p+q\ (\mbox{mod.}\ 2p)$.
More precisely, the relation between $p$ and $q$ determines the one
between $\ell$ and $\ell'$, and each massless mode related with $\ell$
and $\ell'$ is labeled as $n$. The initial conditions are given by
\begin{align}
&U_{n,n+p-q}^{f\prime}
=-\frac{x\langle\phi_0\rangle}{z\langle\psi_{-p}\rangle}
\frac{D_{n,n}^{\alpha,\alpha,S}}{D_{n,n+p-q}^{\alpha,\alpha',S'}}
U_{n,n}^f,\ \ \
U_{n,n}^f=-\frac{x'\langle\phi_0\rangle}{z'\langle\psi_{-p}\rangle}
\frac{D_{n-p-q,n-p-q}^{\alpha',\alpha',S}}
{D_{n-p-q,n}^{\alpha',\alpha,S'}}
U_{n,n-p-q}^{f\prime}.
\label{IC-Type-I}
\end{align}
In that case, the $2p$ massless modes $f_n$ $(n=0,1,\cdots 2p-1)$ can
arise at low energy. 
For the Type-II condition $p=q$, $n$ is defined as $n=\ell=\ell'$.
The initial conditions are given by
\begin{align}
&U_{n,n}^{f\prime}=-\frac{x\langle\phi_0\rangle}{z\langle\psi_{-p}\rangle}
\frac{D_{n,n}^{\alpha,\alpha,S}}{D_{n,n}^{\alpha,\alpha',S'}}
U_{n,n}^f.
\label{IC-Type-II}
\end{align}
The same as for the Type-I condition, the $2p$ massless modes $f_n$
can arise at low energy. 
For the Type-III condition $p<q$, $n$ is defined as
$n=\ell+p-q=\ell'$. The initial conditions are given by
\begin{align}
&U_{n,n}^{f\prime}=-\frac{x\langle\phi_0\rangle}{z\langle\psi_{-p}\rangle}
\frac{D_{n-p+q,n-p+q}^{\alpha,\alpha,S}}{D_{n-p+q,n}^{\alpha,\alpha',S'}}
U_{n,n-p+q}^f.
\label{IC-Type-III}
\end{align}
The same as for the Type-I and -II conditions, the $2p$ massless modes
$f_n$  can arise at low energy. The components
$\hat{f}_{\alpha+k}$ $(k=0,1,\cdots,q-p-1)$ of the matter field
$\hat{F}$ do not contain any massless modes. 

\begin{figure}
\centering
\includegraphics[totalheight=6cm]{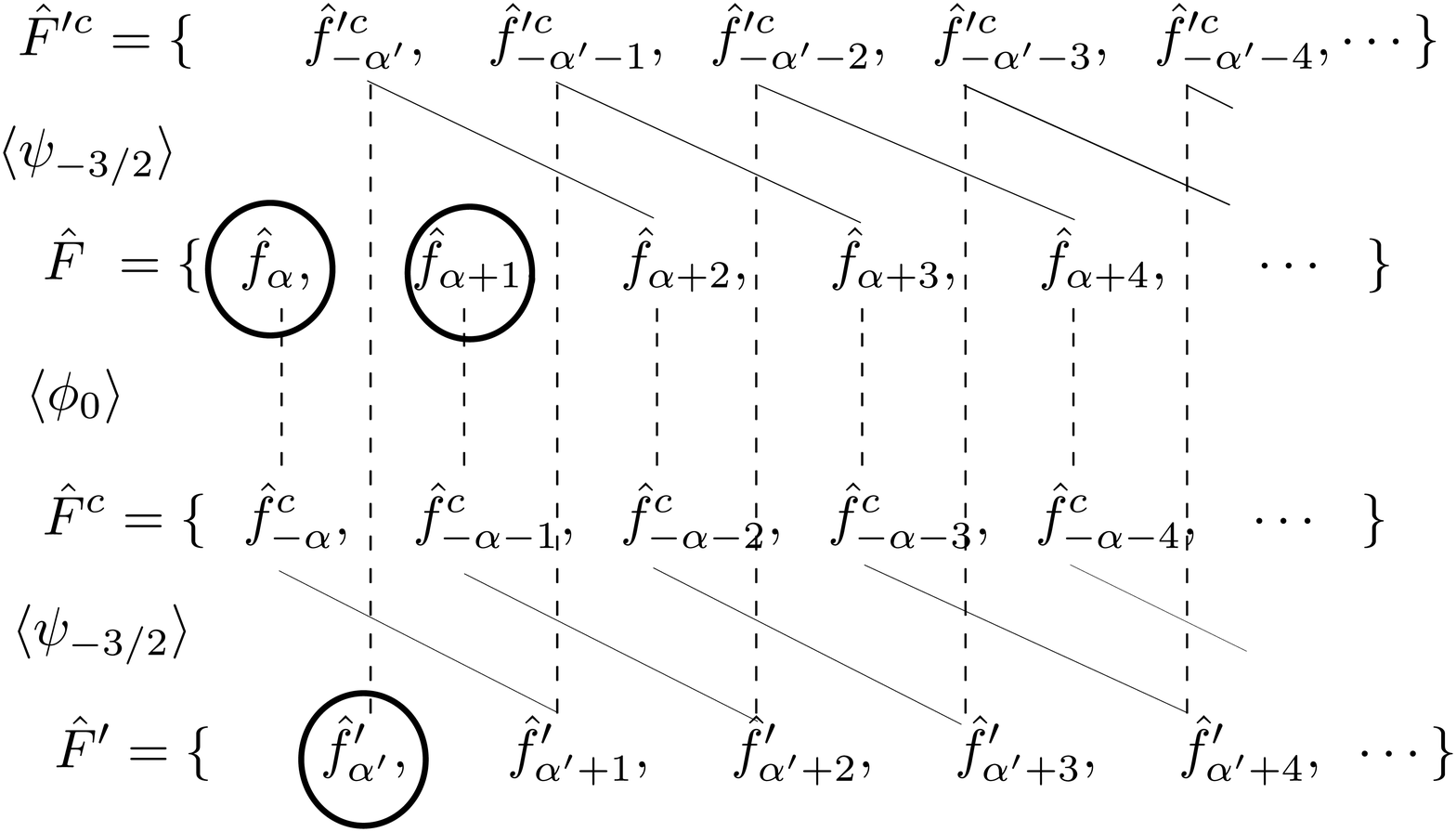}
\caption{An example of Type-I $p=3/2>q=1/2$ for the three massless
generation case in Eq.~(\ref{FF'-mass}):
a massless mode $\hat{f}_0$ is realized by certain linear combinations of
the components $\hat{f}_{\alpha+3k}$ and $\hat{f}'_{\alpha'+1+3k}$
$(k=1,2,\cdots)$, 
and the constructional element of the massless mode is determined  by its
mixing coefficients $U_{0,3k}^f$ and $U_{0,3k+1}^{f\prime}$
given in Eqs.~(\ref{mixing-coefficient-half-1}),
(\ref{mixing-coefficient-half-2}) and (\ref{IC-Type-I}); 
and etc.,
where a main element of each massless chiral mode is surround by
a circle,  
and the components of the matter fields $\hat{F}$, $\hat{F}^c$,
$\hat{F}'$ and $\hat{F}^{\prime c}$
connected by solid and dashed lines have a mass term that 
comes from the VEVs $\langle\psi_{-3/2}\rangle$ and
$\langle\phi_{0}\rangle$, respectively, 
and the mass term of the solid line that comes from the VEV
$\langle\psi_{-3/2}\rangle$ dominantly contributes to whether 
massless modes appear or not.}
\label{fig:SGG-new-p=3/2-T1}
\end{figure}

\begin{figure}
\centering
\includegraphics[totalheight=6cm]{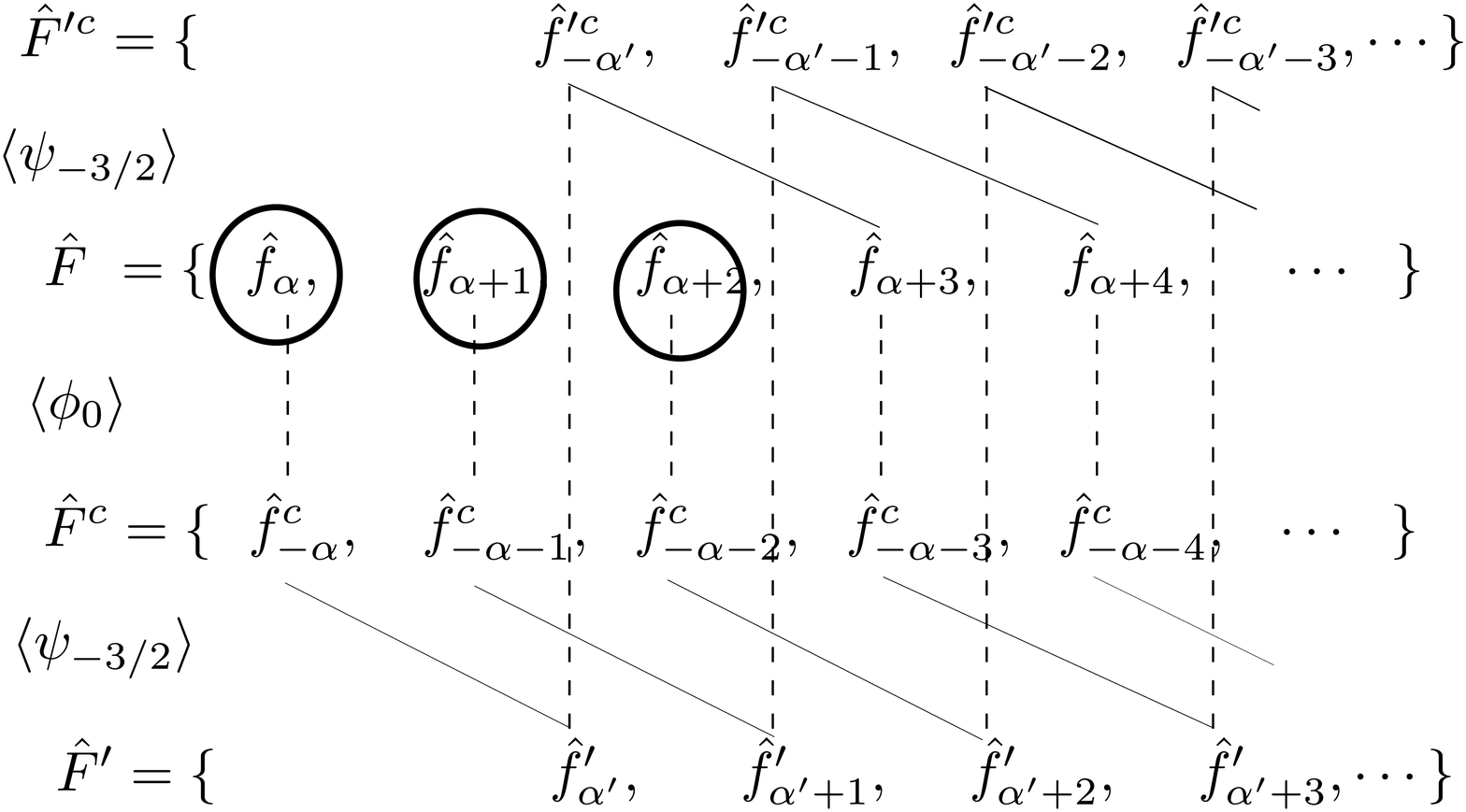}
\caption{An example of Type-II $p=q=3/2$ for the three massless generation
 case in Eq.~(\ref{FF'-mass}):
each massless mode $\hat{f}_n$ is realized by a certain linear combination
of the components $\hat{f}_{\alpha+3k+n}$ and $\hat{f}'_{\alpha'+3k+n}$
$(k=1,2,\cdots)$, 
and the constructional element of its massless mode is determined by its
mixing coefficients $U_{n,3k+n}^f$ and $U_{n,3k+n}^{f\prime}$
given in Eqs.~(\ref{mixing-coefficient-half-1}),
(\ref{mixing-coefficient-half-2}) and (\ref{IC-Type-II}),
where the explanation of the circle and lines is given 
in Fig.~\ref{fig:SGG-new-p=3/2-T1}.}
\label{fig:SGG-new-p=3/2-T2}
\end{figure}

\begin{figure}
\centering
\includegraphics[totalheight=6cm]{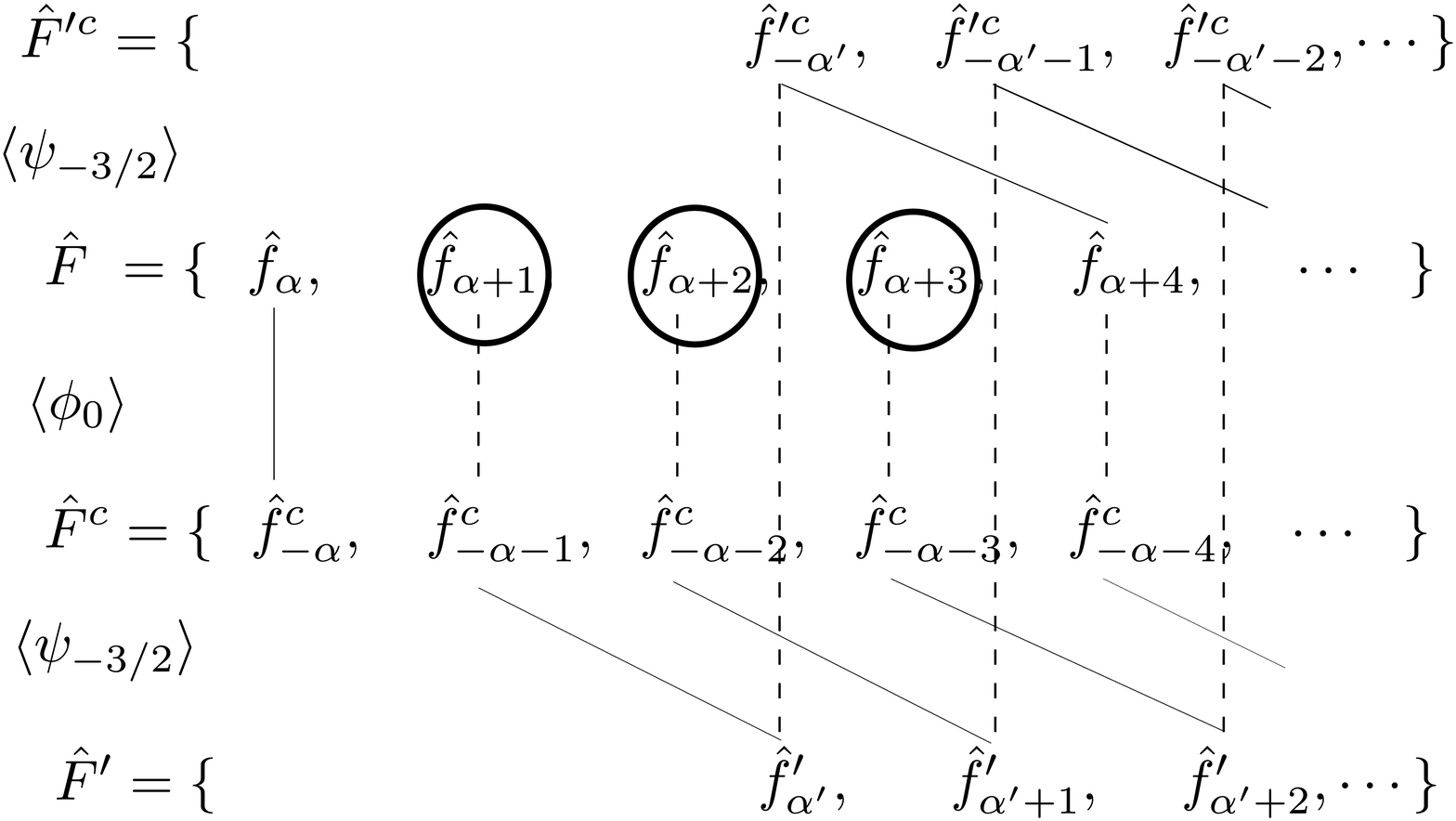}
\caption{An example of Type-III $p=3/2<q=5/2$ for the three massless
generation case in Eq.~(\ref{FF'-mass}):
each massless mode $\hat{f}_n$ is realized by a certain linear combination
of the components $\hat{f}_{\alpha+3k+n+1}$ and $\hat{f}'_{\alpha'+3k+n}$
$(k=1,2,\cdots)$, 
and the constructional element of its massless mode is determined by its
mixing coefficients $U_{n,3k+n+1}^f$ and $U_{n,3k+n}^{f\prime}$
given in Eqs.~(\ref{mixing-coefficient-half-1}),
(\ref{mixing-coefficient-half-2}) and (\ref{IC-Type-III}),
where the explanation of the circle and lines is given 
in Fig.~\ref{fig:SGG-new-p=3/2-T1}.}
\label{fig:SGG-new-p=3/2-T3}
\end{figure}

To clarify the meaning of the above mixing coefficients and the
relation between their initial conditions, we give examples of Type-I
$p=3/2>q=1/2$, Type-II $p=q=3/2$ and Type-III $p=3/2<q=5/2$ 
for the three massless generation case in Eq.~(\ref{FF'-mass}).
First, for $p=3/2$ and $q=1/2$ from Fig.~\ref{fig:SGG-new-p=3/2-T1},
we find that 
the components $\hat{f}_{\alpha}$ and $\hat{f}_{\alpha+1}$ 
of the matter field $\hat{F}$ and the component $\hat{f}'_{\alpha'}$ 
of the matter field $\hat{F}'$ are a main element of massless
modes $\hat{f}_0$, $\hat{f}_1$ and $\hat{f}_2$, respectively;
a massless mode $\hat{f}_0$ is realized by a certain linear combination of
the components $\hat{f}_{\alpha+3k}$ and $\hat{f}'_{\alpha'+1+3k}$
$(k=1,2,\cdots)$, 
and the constructional element of the massless mode is determined  by its
mixing coefficients $U_{0,3k}^f$ and $U_{0,3k+1}^{f\prime}$
given in Eqs.~(\ref{mixing-coefficient-half-1}),
(\ref{mixing-coefficient-half-2}) and (\ref{IC-Type-I}); 
another massless mode $\hat{f}_1$ is realized by a certain linear
combination of the components $\hat{f}_{\alpha+3k+1}$ and
$\hat{f}'_{\alpha'+2+3k}$,
and the element of the massless mode is  determined  by its
mixing coefficients $U_{1,3k+1}^f$ and $U_{1,3k+2}^{f\prime}$; 
and the other massless mode $\hat{f}_2$ is realized by a certain linear
combination of the components 
$\hat{f}'_{\alpha'+3k}$ and $\hat{f}_{\alpha+3k+2}$,
and the element of the massless mode is determined by its
mixing coefficients $U_{2,3k}^{f\prime}$ and $U_{2,3k+2}^f$.
Next, for $p=q=3/2$ from Fig.~\ref{fig:SGG-new-p=3/2-T2},
we find that 
each component $\hat{f}_{\alpha+n}$ $(n=0,1,2)$ of the matter field
$\hat{F}$ is a main element 
of each massless mode $\hat{f}_n$;
each massless mode $\hat{f}_n$ is realized by a certain linear combination
of the components $\hat{f}_{\alpha+3k+n}$ and $\hat{f}'_{\alpha'+3k+n}$
$(k=1,2,\cdots)$, 
and the constructional element of its massless mode is determined by its
mixing coefficients $U_{n,3k+n}^f$ and $U_{n,3k+n}^{f\prime}$
given in Eqs.~(\ref{mixing-coefficient-half-1}),
(\ref{mixing-coefficient-half-2}) and (\ref{IC-Type-II}).
Finally, for $p=3/2$ and $q=5/2$ from Fig.~\ref{fig:SGG-new-p=3/2-T3},
we find that 
each component $\hat{f}_{\alpha+n+1}$ of the matter field $\hat{F}$ is a
main element of each massless mode $\hat{f}_n$;
each massless mode $\hat{f}_n$ is realized by a certain linear combination
of the components $\hat{f}_{\alpha+3k+n+1}$ and $\hat{f}'_{\alpha'+3k+n}$
$(k=1,2,\cdots)$, 
and the constructional element of its massless mode is determined by its
mixing coefficients $U_{n,3k+n+1}^f$ and $U_{n,3k+n}^{f\prime}$
given in Eqs.~(\ref{mixing-coefficient-half-1}),
(\ref{mixing-coefficient-half-2}) and (\ref{IC-Type-III}).
Since the isolated components $\hat{f}_{\alpha}$ and
$\hat{f}_{-\alpha}^c$ have only a mass term from the VEV
$\langle\phi_{0}\rangle$, they are massive.

Regardless of these three types of conditions, the $2p$ massless modes
$f_n$ $(n=0,1,\cdots 2p-1)$ can arise at low energy because the initial
value of $U_{n,\ell}^f$ or $U_{n,\ell'}^{f\prime}$ determines
$U_{n,i}^f$ and $U_{n,j}^{f\prime}$ completely. 
For $p=1/2,3/2,5/2,\cdots,S'$, we can get $1,3,5,\cdots,2S'$ chiral
generations. This means that only odd number generations are allowed.
Especially, if $S'=3/2$ only one or three generations are allowed.

The remaining task is to verify the normalizable condition of the mixing
coefficients $U_{n,i}^f$ and $U_{n,j}^{f\prime}$. Any massless mode
$\hat{f}_n$ $(n=0,1,2,\cdots,2p-1)$ 
must satisfy the normalization conditions 
$\sum_{s=0}^\infty U_{n,\ell+2ps}^f<\infty$ and
$\sum_{s=0}^\infty U_{n,\ell'+2ps}^{f\prime}<\infty$.
This requirement leads to the conditions
\begin{align}
\lim_{i\to\infty}\left|\frac{U_{n,i+2p}^f}{U_{n,i}^f}\right|<1
\ \ \ \mbox{and}\ \ \
\lim_{i\to\infty}\left|\frac{U_{n,i+2p}^{f\prime}}{U_{n,i}^{f\prime}}
\right|<1.
\end{align}
As we discussed before, by using the asymptotic behavior of the CGC
$D_{i,j}^{\alpha,\beta,S}\sim i^S$ given by Eq.~(\ref{D_asymptotic}), 
for $S<S'$, $\sum_{s=0}^\infty|U_{n,\ell+2ps}^f|^2$ and 
$\sum_{s=0}^\infty|U_{n,\ell'+2ps}^{f\prime}|^2$ converge, and then
these massless particles appear at low energy; For $S>S'$, 
$\sum_{k=0}^\infty|U_{n,\ell+2pk}^f|^2$ and
$\sum_{s=0}^\infty|U_{n,\ell'+2ps}^{f\prime}|^2$ diverge, and the
massless ones do not emerge at low energy. This setup cannot satisfy the
marginal assignment $S=S'$ because any integer $S$ is not equal to the
half-integer $S'$. We take a phase convention for the massless modes
$f_n$ so that each larger initial mixing coefficient is real and positive.

\section{Structures of Yukawa Coupling Constants}
\label{sec:Yukawa}

Let us investigate structures of Yukawa coupling constants governed by the
horizontal symmetry $SU(1,1)$. We first introduce three matter fields
$\hat{F}$ and $\hat{G}$ with the lowest weight $+\alpha$ and $+\beta$, 
and $\hat{H}$ with the highest weight $-\gamma$,  
where these weights must satisfy a weight condition
$\Delta:=\gamma-\alpha-\beta$, and $\Delta$ is a non-negative integer
number. Here we consider a Yukawa coupling superpotential term 
\begin{eqnarray}
W_{\rm Yukawa}=y \hat{F}\hat{G}\hat{H},
\label{W-Yukawa-min}
\end{eqnarray}
where $y$ is a dimensionless coupling constant.
We suppose that the matter fields $\hat{F}$ and $\hat{G}$ include three
generations and the matter field $\hat{H}$ includes only one
generation, and the mixing coefficient forms are given by
\begin{equation}
\hat{f}_{\alpha+i}=
\sum_{m=0}^{2} \hat{f}_mU_{m,i}^f+[\cdots],\ \ \ 
\hat{g}_{\beta+i}=
\sum_{m=0}^{2} \hat{g}_mU_{m,i}^g+[\cdots],\ \ \
\hat{h}_{-\gamma-i}=\hat{h}U_i^h+[\cdots],
\label{mixing-coefficients}
\end{equation}
where $m=0,1,2$, and $[\cdots]$ represents the corresponding massive
modes. The explicit expressions for the 
mixing coefficients $U_{m,i}^f$, $U_{m,i}^g$ and $U_i^h$ are determined 
by the mechanism of the spontaneous generation of generations
as we have discussed in \S \ref{sec:SGG}. Since the multiplets
$\hat{F},\hat{G}$ and $\hat{H}$ are unitary representations of
$SU(1,1)$, all coefficients $U$ should be row vectors in unitary
matrices, and then they satisfy 
$\sum_{i=0}^\infty U_{m,i}^{f*}U_{n,i}^f=\delta_{mn}$ and so on. 

The Yukawa coupling constants of the massless mode $\hat{h}$ coupling to 
$\hat{f}_n$ and $\hat{g}_n$ are derived by extracting the massless modes 
from the term in Eq.~(\ref{W-Yukawa-min}) 
\begin{equation}
W_{\rm Yukawa}
=\sum_{m,n=0}^2y^{mn}\hat{f}_m\hat{g}_n\hat{h},
\end{equation}
where $y^{mn}$ are the coupling constants.
By using the CGC in Eq.~(\ref{CGC-C}), the coupling matrix $y^{mn}$ 
is given by
\begin{eqnarray}
y^{mn}=y\sum_{i,j=0}^\infty C_{i,j}^{\alpha,\beta,\Delta}
U_{m,i}^fU_{n,j}^gU_{i+j-\Delta}^h,
\label{y-general}
\end{eqnarray}
where the mixing coefficients $U^{f,g}_{m,i}$ and $U_i^h$
are determined by the $SU(1,1)$ CGCs, the VEVs of structure fields, and
the corresponding coupling constants as we discussed in the previous
section. 

In the rest of this section, we first review the structures of the Yukawa
couplings derived from the mixing coefficients of two structure fields
with an integer spin, which is discussed in
Ref.~\cite{Inoue:2000ia,Yamatsu:2007}. We discuss the Yukawa coupling
structures determined by the mixing coefficients of one structure field
with an integer spin and one with a half-integer spin.

\subsection{Two Structure Fields with an Integer Spin}

Let us examine the structures of Yukawa couplings given by the form of
the mixing coefficients in Eq.~(\ref{mixing-form}) from two structure
fields with integer spins as in Fig.~\ref{fig:SGG-min-g=3-v2}. In this
case, the mixing coefficients are given by 
\begin{align}
&U^f_{n,i}=U^f_n\sum_{s=0}^\infty\delta_{j,n+3s}
(-\epsilon_F)^s b^f_{m,s},\ \
\epsilon_F=
\frac{x_F\langle\psi_{F0}\rangle}{x_F'\langle\psi_{F-3}'\rangle},\ \ 
b_{n,s}^f=\prod_{r=0}^{s-1}
\frac{D^{\alpha,\alpha,S_F}_{3r+n,3r+n}}
{D^{\alpha,\alpha,S_F'}_{3r+n,3(r+1)+n}},
\label{minimal-mixing-f}\\
&U^g_{n,i}=U^g_n\sum_{s=0}^\infty\delta_{i,n+3s}
(-\epsilon_G)^s b^g_{m,s},\ \
\epsilon_G=
\frac{x_G\langle\psi_{G0}\rangle}{x_G'\langle\psi_{G-3}'\rangle},\ \
b_{n,s}^g=\prod_{r=0}^{s-1}
\frac{D^{\beta,\beta,S_G}_{3r+n,3r+n}}
{D^{\beta,\beta,S_G'}_{3r+n,3(r+1)+n}},
\label{minimal-mixing-g}\\
&U_{i}^h=U_0^h(-\epsilon_H)^i b_{i}^h,\ \
\epsilon_H=
\frac{x_H\langle\psi_{H0}\rangle}{x_H'\langle\psi_{H+1}'\rangle},\ \
b_{i}^h=\prod_{r=0}^{i-1}
\frac{D^{\gamma,\gamma,S_H}_{r,r}}{D^{\gamma,\gamma,S_H'}_{r,r+1}},
\label{minimal-mixing-h}
\end{align}
where the subscripts $F,G,H$ and the superscripts $f,g,h$ represent 
the matter fields $\hat{F},\hat{G},\hat{H}$.
From the mixing coefficients
in Eqs.~(\ref{minimal-mixing-f})-(\ref{minimal-mixing-h}), the Yukawa
coupling in Eq.~(\ref{y-general}) is in the form
\begin{equation}
y^{mn}=yU^f_mU^g_nU_0^h\sum_{r,s=0}^\infty 
(\epsilon_F\epsilon_H^3)^{r}(\epsilon_G\epsilon_H^3)^{s}
(-\epsilon_H)^{m+n-\Delta}
C_{m+3r,n+3s}^{\alpha,\beta,\Delta}
b^f_{m,r}b^g_{n,s}b^h_{m+n-\Delta+3(r+s)},
\label{y-general-minimal-general}
\end{equation}
where $\Delta:=\gamma-\alpha-\beta$.

We consider the contribution of the nonzero $s$ and/or $r$ terms
for the Yukawa coupling in Eq.~(\ref{y-general-minimal-general}).
The asymptotic behavior of
$D_{j,i}^{\beta,\alpha,S}$ for the large $S$ limit with $|i-j|$ fixed,
\begin{align}
D_{j,i}^{\beta,\alpha,S}\simeq
(-i)^iS^{i+j+\alpha+\beta-1}
\frac{1}{\sqrt{i!j!\Gamma(2\alpha+i)\Gamma(2\beta+j)}},
\end{align}
leads to 
\begin{align}
&b_{n,s}^f\simeq
\prod_{r=0}^{s-1}
\frac{(S_F)^{6r+2n+2\alpha-1}}{(S_F')^{6r+2n+2\alpha+2}}
\sqrt{\frac{(3r+n+3)!\Gamma(2\alpha+3r+n+3)}
{(3r+n)!\Gamma(2\alpha+3r+n)}},\nonumber\\
&b_{i}^h\simeq
\prod_{r=0}^{i-1}
\frac{(S_H)^{2r+2\gamma-1}}{(S_H')^{2r+2\gamma}}\sqrt{(r+1)(2\gamma+r)},
\end{align}
where $b_{n,s}^g$ is given by replacing $F$ and $\alpha$ in $b_{n,s}^f$
by $G$ and $\beta$.
Note that we only consider the cases with $S_F\leq S_F'$, $S_G\leq S_G'$
and $S_H\leq S_H'$ because each massless mode is absent in the cases with 
$S_F>S_F'$, $S_G>S_G'$, and $S_H>S_H'$, and there are no Yukawa couplings
between massless modes.
The $b_{n,s}^f$ quickly decreases as the $SU(1,1)$ spin becomes large,
so the contribution from the nonzero $s$ and/or $r$ terms is negligible.
Next we consider cases with smaller $SU(1,1)$ spins, where
$S_F'$ and $S_G'$ must be larger than or equal to three, and $S_H'$ must
be larger than or equal to one due to the existence of the nonvanishing
VEVs of the structure fields. For $S_F<S_F'$, $S_G<S_G'$ and $S_H<S_H'$,
typical weights $\alpha$, $\beta$ and $\gamma$ give
$b_{n,0}^f\gg b_{n,1}^f\gg\cdots$, $b_{n,0}^g\gg b_{n,1}^g\gg\cdots$ and 
$b_{n}^h\gg b_{n+3}^h\gg\cdots$, respectively. Thus,
even if $|\epsilon_F|$ and $|\epsilon_G|\simeq 1$, 
the contribution of the terms $(r+s>0)$ in 
Eq.~(\ref{y-general-minimal-general}) is negligible.
For $S_F=S_F'$, $S_G=S_G'$ and $S_H=S_H'$, 
$b_{n,0}^f\simeq b_{n,1}^f\simeq\cdots$, 
$b_{n,0}^g\simeq b_{n,1}^g\simeq\cdots$ and  
$b_{n}^h\simeq b_{n+3}^h\simeq\cdots$, respectively, but 
$C_{i,j}^{\alpha,\beta,\Delta}>C_{i+3,j}^{\alpha,\beta,\Delta}\simeq
C_{i,j+3}^{\alpha,\beta,\Delta}>C_{i+3,j+3}^{\alpha,\beta,\Delta}>\cdots$
and the parameters $\epsilon$ must satisfy the relations
$\epsilon_{F,G}<\epsilon_{F,Gcr}^{g=3}$ and
$\epsilon_{H}<\epsilon_{Hcr}^{g=1}$ to produce each corresponding
massless mode, where 
$\left.1/\sqrt{20}\right|_{S_{F,G}=S_{F,G}'=3}\leq\epsilon_{F,Gcr}^{g=3}
<\left.1\right|_{S_{F,G}=S_{F,G}'\to\infty}$ and 
$\left.1/\sqrt{2}\right|_{S_H=S_H'=1}\leq\epsilon_{Hcr}^{g=1}<
\left.1\right|_{S_H=S_H'\to\infty}$
from Eq.~(\ref{epsilon-critical}). Thus, we can neglect the
contributions from the nonzero $s$ and/or $r$ terms in
Eq.~(\ref{y-general-minimal-general}) as long as the corresponding
$s=r=0$ term exists. 
That is we can use the mixing coefficients of three generations in 
Fig.~\ref{fig:SGG-min-g=3} instead of the ones in 
Fig.~\ref{fig:SGG-min-g=3-v2}.
When $\Delta>0$, some matrix elements have no contribution from the
$s=r=0$ term. For example, for $\Delta=1$, the $(r,s)=0$ term of
$y^{00}$ is not allowed and the $(r,s)=(1,0),(0,1)$ terms dominantly
contribute to this matrix. 
In the following discussion, we analyze Yukawa matrices for the
$s=r=0$ term in Eq.~(\ref{y-general-minimal-general}) 
\begin{equation}
y^{mn}=yU_m^fU_n^gU_0^h
(-\epsilon_H)^{m+n-\Delta}C_{m,n}^{\alpha,\beta,\Delta}
b^h_{m+n-\Delta}.
\label{y-minimal}
\end{equation}

Let us examine this structure for large $SU(1,1)$ spin $S_H'$. In this
case, since $b_i^h\simeq 0$ for $(i\geq 1)$,
$U^h_i\simeq\delta_{i,0}$. Thus, the massless mode $\hat{h}$ emerges as 
the pure $0$th component of $\hat{h}_{-\gamma-i}$.
For $\Delta\geq 0$, the Yukawa coupling matrix is 
$y^{mn}=yC_{m,n}^{\alpha,\beta,\Delta}\delta_{m+n-\Delta,0}$.
For $\Delta=0,4$, the rank of this matrix is one, and only one
generation of each of $\hat{f}_m$ and $\hat{g}_n$ has a mass when
$\hat{h}$ has a nonvanishing VEV.
For $\Delta=1,3$, the rank of this matrix is two, and two generations of
$\hat{f}_m$ and $\hat{g}_n$ have degenerate masses.
For $\Delta=2$, the rank of this matrix is three, and three generations
of $\hat{f}_m$ and $\hat{g}_n$ have almost degenerate masses.
For $\Delta\geq 5$, there is no Yukawa coupling. 

Let us begin to analyze the structure of the Yukawa couplings in
Eq.~(\ref{y-minimal}) for small $S_H'$. We here concentrate our
discussion on the $\Delta=0$ case. To examine this pattern, it is
convenient to introduce the normalized matrix
$\tilde{y}^{mn}:=y^{mn}/y^{00}$.
By using the CGC in Eq.~(\ref{CGC-C}), the normalized matrix
$\tilde{y}^{mn}$ is given by  
\begin{equation}
\tilde{y}^{mn}:=\frac{y^{mn}}{y^{00}}
=\tilde{U}_{m}^{f}\tilde{U}_{n}^{g}
\epsilon_H^{m+n}\sqrt{\frac{(m+n)!}{m!n!}
\frac{\Gamma(2\gamma)\Gamma(2\alpha+m)\Gamma(2\beta+n)}
{\Gamma(2\gamma+m+n)\Gamma(2\alpha)\Gamma(2\beta)}}\ b_{m+n}^{h},
\label{normalized_Yukawa}
\end{equation}
where $b^{h}_0=1$, $\tilde{U}^{f}_{m}:=U^{f}_{m}/U^{f}_{0}$,
and $\tilde{U}^{g}_{n}:=U^{g}_{n}/U^{g}_{0}$.
The explicit form is
\begin{eqnarray}
\tilde{y}^{mn}=
\left(\begin{array}{lll}
1
&\tilde{U}_{1}^{g}\epsilon_H
\sqrt{\frac{\beta}{\gamma}} b^{h}_1
&\tilde{U}_{2}^{g}\epsilon_H^2
\sqrt{\frac{(2\beta+1)\beta}{(2\gamma+1)\gamma}} b^{h}_2\\
\tilde{U}_{1}^{f}\epsilon_H
\sqrt{\frac{\alpha}{\gamma}} b^{h}_1
&\tilde{U}_{1}^{f}\tilde{U}_{1}^{g}\epsilon^2_H
\sqrt{\frac{4\alpha\beta}{(2\gamma+1)\gamma}} b^{h}_2
&\tilde{U}_{1}^{f}\tilde{U}_{2}^{g}\epsilon_H^3
\sqrt{\frac{3(2\beta+1)\alpha\beta}{\gamma(\gamma+1)(2\gamma+1)}}
b^{h}_3\\ 
\tilde{U}_{2}^{f}\epsilon_H^2
\sqrt{\frac{(2\alpha+1)\alpha}{(2\gamma+1)\gamma}} b^{h}_2
&\tilde{U}_{2}^{f}\tilde{U}_{1}^{g}\epsilon_H^3
\sqrt{\frac{3(2\alpha+1)\alpha\beta}{\gamma(\gamma+1)(2\gamma+1)}}
b^{h}_3
&\tilde{U}_{2}^{f}\tilde{U}_{2}^{g}\epsilon_H^4
\sqrt{\frac{6(2\alpha+1)(2\beta+1)\alpha\beta}{\gamma(\gamma+1)
(2\gamma+1)(2\gamma+3)}} b^{h}_4
\end{array}
\right).
\end{eqnarray}
This matrix is diagonalized by a bi-unitary transformation because 
it is a complex matrix:
\begin{align}
V^\dag \tilde{y}^{mn} U=\mbox{diag}(\tilde{y}_0,\tilde{y}_1,\tilde{y}_2),
\end{align}
where $V$ and $U$ are unitary matrices.

To understand the features of the Yukawa coupling matrix in
Eq.~(\ref{normalized_Yukawa}), we expand the eigenvalues
$\tilde{y}_0,\tilde{y}_1$,$\tilde{y}_2$ of this type of matrix 
in a power series in $\epsilon_H^2$ by
$\tilde{y}_0=w_0+O(\epsilon_H^2)$,
$\tilde{y}_1=\epsilon^2w_1+O(\epsilon_H^4)$, and
$\tilde{y}_2=\epsilon^4w_2+O(\epsilon_H^6)$.
The leading terms have the expressions
$w_0=\tilde{y}^{00}$,
$\epsilon^2w_1=\mbox{det}(\tilde{y}^{mn})_{2\times 2}/\tilde{y}^{00}$,
and 
$\epsilon^4w_2
=\mbox{det}(\tilde{y}^{mn})_{3\times 3}/
\mbox{det}(\tilde{y}^{mn})_{2\times 2}$,
where 
$\mbox{det}(\tilde{y}^{mn})_{2\times 2}:=
\tilde{y}^{00}\tilde{y}^{11}-\tilde{y}^{01}\tilde{y}^{10}$, and 
$\mbox{det}(\tilde{y}^{mn})_{3\times 3}:=
\tilde{y}^{22}(\tilde{y}^{00}\tilde{y}^{11}-\tilde{y}^{01}\tilde{y}^{10})
-\tilde{y}^{21}(\tilde{y}^{00}\tilde{y}^{12}-\tilde{y}^{10}\tilde{y}^{02})-\tilde{y}^{20}(\tilde{y}^{11}\tilde{y}^{02}-\tilde{y}^{01}\tilde{y}^{12})$.
The straightforward calculation gives the expressions of
$\tilde{y}_0,\tilde{y}_1,\tilde{y}_2$ in the forms
\begin{align}
&\tilde{y}_0=1+O(\epsilon_H^2),\nonumber\\
&\tilde{y}_1=\epsilon_H^2
\sqrt{\frac{\alpha\beta}{(2\gamma+1)\gamma}}
\tilde{U}^{f}_{1}\tilde{U}^{g}_{1}
\ X_\gamma
+O(\epsilon_H^4),\nonumber\\
&\tilde{y}_2=\epsilon_H^4
\sqrt{\frac{6\alpha\beta(2\alpha+1)(2\beta+1)}
{\gamma(\gamma+1)(2\gamma+1)(2\gamma+3)}}
\tilde{U}^{f}_{2}\tilde{U}^{g}_{2}
\ \frac{Z_\gamma}{X_\gamma}
+O(\epsilon_H^6),
\label{normalized_Yukawa_solutions}
\end{align}
where the $X_\gamma$ and $Z_\gamma$ only depend on the representations of
$SU(1,1)$: 
\begin{align}
X_\gamma:=&2b_2^{h}-b_1^{h2}\sqrt{\frac{2\gamma+1}{\gamma}},\\
Z_\gamma:=&2b_2^{h}b_4^{h}
+2b_1^{h}b_2^{h}b_3^{h}\sqrt{\frac{2\gamma+3}{2\gamma}}
-3b_3^{h2}\sqrt{\frac{2\gamma+3}{6(\gamma+1)}}
\nonumber\\&
-2b_2^{h3}\sqrt{\frac{(\gamma+1)(2\gamma+3)}{6\gamma(2\gamma+1)}}
-b_1^{h2}b_4^{h}\sqrt{\frac{2\gamma+1}{\gamma}}.
\end{align}
In this case, the unitary matrices $U$ and $V$ are approximately
expressed by 
{\small 
\begin{eqnarray}
&&\hspace{-2em}
U\simeq\left(\begin{array}{ccc}
1
&-\epsilon_H\tilde{U}^{g}_{1}
\sqrt{\frac{\beta}{\gamma}}b^{h}_1
&-\epsilon_H^2\tilde{U}^{g}_{2}
\frac{\sqrt{\frac{4\beta(2\beta+1)}{2\gamma+1}}b^{h2}_2
-\sqrt{\frac{3\beta(2\beta+1)}{\gamma+1}}b^{h}_1b^{h}_3}
{2\sqrt{\gamma}b^{h}_2-\sqrt{2\gamma+1}b^{h2}_1}\\
\epsilon_H\tilde{U}^{g}_{1}
\sqrt{\frac{\beta}{\gamma}}b^{h}_1
&1
&-\epsilon_H\frac{\tilde{U}^{g}_{2}}{\tilde{U}^{g}_{1}}
\frac{\sqrt{\frac{3(2\beta+1)\gamma}{\gamma+1}}b^{h}_3
-\sqrt{2\beta+1}b^{h}_1b^{h}_2}
{2\sqrt{\gamma}b^{h}_2-\sqrt{2\gamma+1}b^{h2}_1}\\
\epsilon_H^2\tilde{U}^{g}_{2}
\sqrt{\frac{(2\beta+1)\beta}{(2\gamma+1)\gamma}}b^{h}_2
&\epsilon_H\frac{\tilde{U}^{g}_{2}}{\tilde{U}^{g}_{1}}
\frac{\sqrt{\frac{3(2\beta+1)\gamma}{\gamma+1}}b^{h}_3
-\sqrt{2\beta+1}b^{h}_1b^{h}_2}
{2\sqrt{\gamma}b^{h}_2-\sqrt{2\gamma+1}b^{h2}_1}
&1\\
\end{array}
\right),\nonumber\\
\\
&&\hspace{-2em}
V\simeq\left(\begin{array}{ccc}
1
&-\epsilon_H\tilde{U}^{f}_{1}
\sqrt{\frac{\alpha}{\gamma}}b^{h}_1
&-\epsilon_H^2\tilde{U}^{f}_{2}
\frac{\sqrt{\frac{4\alpha(2\alpha+1)}{2\gamma+1}}b^{h2}_2
-\sqrt{\frac{3\alpha(2\alpha+1)}{\gamma+1}}b^{h}_1b^{h}_3}
{2\sqrt{\gamma}b^{h}_2-\sqrt{2\gamma+1}b^{h2}_1}\\
\epsilon_H\tilde{U}^{f}_{1}\sqrt{\frac{\alpha}{\gamma}}b^{h}_1
&1
&-\epsilon_H
\frac{\tilde{U}^{f}_{2}}{\tilde{U}^{f}_{1}}
\frac{\sqrt{\frac{3(2\alpha+1)\gamma}{\gamma+1}}b^{h}_3
-\sqrt{2\alpha+1}b^{h}_1b^{h}_2}
{2\sqrt{\gamma}b^{h}_2-\sqrt{2\gamma+1}b^{h2}_1}\\
\epsilon_H^2\tilde{U}^{f}_{2}
\sqrt{\frac{(2\alpha+1)\alpha}{(2\gamma+1)\gamma}}b^{h}_2
&\epsilon_H
\frac{\tilde{U}^{f}_{2}}{\tilde{U}^{f}_{1}}
\frac{\sqrt{\frac{3(2\alpha+1)\gamma}{\gamma+1}}b^{h}_3
-\sqrt{2\alpha+1}b^{h}_1b^{h}_2}
{2\sqrt{\gamma}b^{h}_2-\sqrt{2\gamma+1}b^{h2}_1}
&1\\
\end{array}
\right).\nonumber\\
\end{eqnarray}}

To understand the hierarchical pattern, it is useful to see some
examples. A set of $SU(1,1)$ spins $(S_H,S_H')=(0,1)$ determines the
$b_i^h$ in Eq.~(\ref{minimal-mixing-h}) given by
\begin{align}
b_i^h=\sqrt{\frac{\Gamma(2\gamma)}{2^ii!\Gamma(2\gamma+i)}}.
\end{align}
This leads to
\begin{align}
&\tilde{y}_0=1+O(\epsilon_H^2),\nonumber\\
&\tilde{y}_1=
-\epsilon_H^2\frac{\sqrt{\alpha\beta}}{2\gamma^2(2\gamma+1)}+
O(\epsilon_H^4),\nonumber\\ 
&\tilde{y}_2=\epsilon_H^4
\frac{\sqrt{\alpha\beta(2\alpha+1)(2\beta+1)}}
{4\gamma^3(\gamma+1)^2(2\gamma+1)^2(2\gamma+3)}+O(\epsilon_H^6).
\end{align}
We can easily calculate the eigenvalues for some weight sets:
For the weight set $\alpha=\beta=\gamma/2=1/2$,
$\tilde{y}_0=1+O(\epsilon_H^2)$,
$\tilde{y}_1=-\epsilon_H^2/12+O(\epsilon_H^4)$,
and $\tilde{y}_2=\epsilon_H^4/720+O(\epsilon_H^6)$.
For $\alpha=\beta=\gamma/2=1$,
$\tilde{y}_0=1+O(\epsilon_H^2)$,
$\tilde{y}_1=-\epsilon_H^2/40+O(\epsilon_H^4)$, and 
$\tilde{y}_2=\epsilon_H^4/16800+O(\epsilon_H^6)$.
For $\alpha=\beta=\gamma/2=2$,
$\tilde{y}_0=1+O(\epsilon_H^2)$,
$\tilde{y}_1=-\epsilon_H^2/144+O(\epsilon_H^4)$, and 
$\tilde{y}_2=\epsilon_H^4/570240+O(\epsilon_H^6)$.
For $\alpha=\beta=\gamma/2\to\infty$,
$\tilde{y}_0\simeq 1$, $\tilde{y}_1\simeq 0$, and $\tilde{y}_2\simeq 0$.
Another set of $SU(1,1)$ spins $(S_H,S_H')=(1,1)$ provides the $b_i^h$
given by 
\begin{align}
b_i^h=\sqrt{\frac{2^i\Gamma(2\gamma)}{i!\Gamma(2\gamma+i)}}
\frac{\Gamma(\gamma+i)}{\Gamma(\gamma)}
\end{align}
and this gives 
\begin{align}
&\tilde{y}_0=1+O(\epsilon_H^2),\nonumber\\
&\tilde{y}_1=-\epsilon_H^2\frac{\sqrt{\alpha\beta}}{2(2\gamma+1)}
+O(\epsilon_H^4),\nonumber\\ 
&\tilde{y}_2=\epsilon_H^4
\frac{\gamma\sqrt{\alpha\beta(2\alpha+1)(2\beta+1)}}
{4(2\gamma+1)^2(2\gamma+3)}
+O(\epsilon_H^6).
\end{align}
We also give the eigenvalues for some weight sets:
For the weight set $\alpha=\beta=\gamma/2=1/2$,
$\tilde{y}_0=1+O(\epsilon_H^2)$,
$\tilde{y}_1=\epsilon_H^2/12+O(\epsilon_H^4)$, and 
$\tilde{y}_2=\epsilon_H^4/180+O(\epsilon_H^6)$.
For $\alpha=\beta=\gamma/2=1$,
$\tilde{y}_0=1+O(\epsilon_H^2)$,
$\tilde{y}_1=\epsilon_H^2/10+O(\epsilon_H^4)$, and 
$\tilde{y}_2=3\epsilon_H^4/700+O(\epsilon_H^6)$.
For $\alpha=\beta=\gamma/2=2$,
$\tilde{y}_0=1+O(\epsilon_H^2)$,
$\tilde{y}_1=\epsilon_H^2/9+O(\epsilon_H^4)$, and 
$\tilde{y}_2=10\epsilon_H^4/891+O(\epsilon_H^6)$.
For $\alpha=\beta=\gamma/2\to\infty$,
$\tilde{y}_0=1+O(\epsilon_H^2)$,
$\tilde{y}_1=\epsilon_H^2/8+O(\epsilon_H^4)$, and 
$\tilde{y}_2=\epsilon_H^4/64+O(\epsilon_H^6)$.
From the above, we determine that for the marginal assignment
$(S_H=S_H')$ the eigenvalues of the Yukawa coupling constants are
relatively stable with respect to  changing the weights, but for the
other assignments $(S_H<S_H')$ they are very sensitive.

We now summarize the results of this part. 
For $S_H<S_H'$, as the weight $\gamma$ increases, the
eigenvalues $\tilde{y}_1$ and $\tilde{y}_2$ decrease. 
For $\gamma\to\infty$, $\tilde{y}_1$ and $\tilde{y}_2$ become
zero. For the marginal assignment $S_H=S_H'$, $\tilde{y}_1$ and
$\tilde{y}_2$ become 
certain non-zero values, which depend on the value of the $SU(1,1)$ spin
and the weights $\alpha$ and $\beta$. Even in this case, for large
$SU(1,1)$ spin $S_H(=S_H')$, $y_1$ and $y_2$ become zero. 
The mass parameters at the GUT scale in the MSSM are given by
Ref.~\cite{Ross:2007az} for several values of $\tan\beta$ by using 
the renormalization group equations of the two-loop gauge couplings and
the two-loop Yukawa couplings assuming an effective SUSY scale of 500
GeV. 
Translating these mass parameters for $\tan\beta=10$ into coupling
constants, the constants for the up-type quark,
the down-type quark, and the charged lepton are 
$(\tilde{y}_0,\tilde{y}_1,\tilde{y}_2)
\simeq(1,2.5\times 10^{-3},6.7\times 10^{-6})$,
$(1,2\times 10^{-2},1.0\times 10^{-3})$, and 
$(1,6\times 10^{-2},2.5\times 10^{-4})$, respectively,
and the values are almost the same for $\tan\beta=38$ and $50$.
When we take the value of the $\epsilon_H$ to range from
$0.1$ to $1$, small $SU(1,1)$ spins are preferred and 
small $SU(1,1)$ weights are also preferred except for the marginal
assignment.

In this paper, we will not discuss the CKM
\cite{Cabibbo:1963yz,Kobayashi:1973fv} and MNS matrices \cite{Maki:1962mu}
because these matrices are very dependent on the choice of the weights
and spins of $SU(1,1)$, where of course we need to set up the neutrino
sector. Some discussions of the CKM and the MNS matrices in this model
can be found in Ref.~\cite{Inoue:2000ia} and \cite{Inoue:2003qi},
respectively. 

\subsection{One Structure Field with an Integer Spin and One with a Half-Integer Spin}

We introduce additional matter fields 
$\hat{F}'$ and $\hat{G}'$ with the lowest weights $\alpha'$ and $\beta'$ 
and $\hat{H}'$ with the highest weight $-\gamma'$, 
which have the same representations
as the matter fields $\hat{F}$, $\hat{G}$, $\hat{H}$ except that of
$SU(1,1)$, respectively.  
In this case, some weight conditions allow not only one Yukawa coupling 
superpotential term but also another. This corresponds to the
extended higgs sector discussed in Ref.~\cite{Yamatsu:2007}. One of the
main aims of this paper is to show the effects of the structure fields
with half-integer spins of $SU(1,1)$. Thus, in this paper we will not
discuss the multi-Yukawa coupling constants and we only discuss some
examples of the structures of each Yukawa coupling constant. 

We suppose that the matter fields $\hat{F}+\hat{F}'$ and
$\hat{G}+\hat{G}'$ include three generations and the matter fields
$\hat{H}+\hat{H}'$ include only one generation and the following mixing
coefficient forms are 
\begin{align}
&\hat{f}_{\alpha+i}=
\sum_{m=0}^{2} \hat{f}_mU_{m,i}^f+[\cdots],\ \ \ 
\hat{g}_{\beta+i}=
\sum_{m=0}^{2} \hat{g}_mU_{m,i}^g+[\cdots],\ \ \
\hat{h}_{-\gamma-i}=\hat{h}U_i^h+[\cdots],\nonumber\\
&\hat{f}_{\alpha'+i}'=
\sum_{m=0}^{2} \hat{f}_m'U_{m,i}^{f\prime}+[\cdots],\ \ \ 
\hat{g}_{\beta'+i}'=
\sum_{m=0}^{2} \hat{g}_m'U_{m,i}^{g\prime}+[\cdots],\ \ \
\hat{h}_{-\gamma'-i}'=\hat{h}'U_i^{h\prime}+[\cdots],
\label{mixing-coefficients-2}
\end{align}
where $q_\alpha:=\alpha'-\alpha,q_\beta:=\beta'-\beta,
q_\gamma:=\gamma'-\gamma$, and  the
$q$s are positive and half-integer numbers.

As before, we consider a superpotential term of Yukawa
couplings  
\begin{align}
W=y\hat{F}^{(\prime)}\hat{G}^{(\prime)}\hat{H}^{(\prime)}
=y\sum_{i,j=0}^\infty
C_{i,j}^{\alpha^{(\prime)},\beta^{(\prime)},\Delta} 
\hat{f}^{(\prime)}_{\alpha^{(\prime)}+i}
\hat{g}^{(\prime)}_{\beta^{(\prime)}+j}
\hat{h}^{(\prime)}_{-\gamma^{(\prime)}-i-j+\Delta},
\end{align}
where $\Delta:=\gamma^{(\prime)}-\alpha^{(\prime)}-\beta^{(\prime)}$ and
$\Delta$ is a semi-positive integer.
$\hat{F}^{(\prime)}\hat{G}^{(\prime)}\hat{H}^{(\prime)}$
stands for every combination of 
$\hat{F}^{(\prime)}$, $\hat{G}^{(\prime)}$ and $\hat{H}^{(\prime)}$:
i.e., 
$\hat{F}\hat{G}\hat{H}$,
$\hat{F}\hat{G}\hat{H}^{\prime}$,
$\hat{F}\hat{G}^{\prime}\hat{H}$,
$\hat{F}^{\prime}\hat{G}\hat{H}$,
$\hat{F}\hat{G}^{\prime}\hat{H}^{\prime}$,
$\hat{F}^{\prime}\hat{G}\hat{H}^{\prime}$,
$\hat{F}^{\prime}\hat{G}^{\prime}\hat{H}$ and 
$\hat{F}^{\prime}\hat{G}^{\prime}\hat{H}^{\prime}$.
By using the mixing coefficients in Eq.~(\ref{mixing-coefficients-2}), we
determine the Yukawa coupling constants
\begin{align}
y^{mn}=y\sum_{i,j=0}^\infty 
C_{i,j}^{\alpha^{(\prime)},\beta^{(\prime)},\Delta}
U_{m,i}^{f(\prime)}U_{n,j}^{g(\prime)}U_{i+j-\Delta}^{h(\prime)},
\end{align}
where $U_{i+j-\Delta}^{h(\prime)}=0$ for $i+j<\Delta$.
When the massless modes are produced by the superpotential
in Eq.~(\ref{Wf-min}) and the mixing coefficients are of the form in
Eqs.~(\ref{mixing-coefficient-half-1}) and
(\ref{mixing-coefficient-half-2}) with certain initial conditions
in Eqs.~(\ref{IC-Type-I})-(\ref{IC-Type-III}), 
the Yukawa coupling matrix is 
\begin{align}
y^{mn}
&=yU_{m,\ell^{(\prime)}}^{f(\prime)}U_{n,k^{(\prime)}}^{g(\prime)}
U_{v^{(\prime)}}^{h(\prime)}
\sum_{r,s=0}^\infty 
(\epsilon_F\epsilon_H^3)^r(\epsilon_G\epsilon_H^3)^s
\epsilon_H^{\ell^{(\prime)}+k^{(\prime)}-v^{(\prime)}-\Delta}
\nonumber\\
&\hspace{2em}\times
C_{\ell^{(\prime)}+3r,k^{(\prime)}+3s}^{\alpha^{(\prime)},\beta^{(\prime)},\Delta}
b_{\ell^{(\prime)},r}^{f(\prime)}b_{k^{(\prime)},s}^{g(\prime)}
b_{\ell^{(\prime)}+k^{(\prime)}-v^{(\prime)}+3(r+s)-\Delta}^{h(\prime)},
\label{y-half-spins}
\end{align}
where $\ell^{(\prime)}$, $k^{(\prime)}$ and $v^{(\prime)}$ represent 
the components of each massless mode of $\hat{F}^{(\prime)}$,
$\hat{G}^{(\prime)}$ and 
$\hat{H}^{(\prime)}$, respectively, and 
$U_{m,\ell^{(\prime)}}^{f(\prime)}$,
$U_{n,k^{(\prime)}}^{g(\prime)}$, and 
$U_{v^{(\prime)}}^{h(\prime)}$ are each the initial condition of
$U_{m,i}^{f(\prime)}$, $U_{n,j}^{g(\prime)}$, and 
$U_{i+j-\Delta}^{h(\prime)}$, respectively.

We analyze the Yukawa coupling in the case $\Delta=0$ and the
Type-II of $\hat{H}^{(\prime)}$ ($v^{(\prime)}=0$)
as in Fig.~\ref{fig:SGG-new-p=3/2-T2}. 
The mixing effect from the non-zero $r$ and/or $s$ terms in
Eq.~(\ref{y-half-spins}) is neglected as before. The Yukawa coupling
constant is 
\begin{align}
y^{mn}
&=yU_{m,\ell^{(\prime)}}^{f(\prime)}U_{n,k^{(\prime)}}^{g(\prime)}
U_{0}^{h(\prime)}
\epsilon_H^{\ell^{(\prime)}+k^{(\prime)}}
C_{\ell^{(\prime)},k^{(\prime)}}^{\alpha^{(\prime)},\beta^{(\prime)},0}
b_{\ell^{(\prime)}+k^{(\prime)}}^{h(\prime)},
\end{align}
where we have used $b_{\ell,0}^{f(\prime)}=b_{\ell',0}^{g(\prime)}=1$.
The normalized Yukawa coupling can be defined in the same way as in
Eq.~(\ref{normalized_Yukawa}), and we obtain the solutions 
Eq.~(\ref{normalized_Yukawa_solutions}) except 
Type-I $\hat{F}'$ and $\hat{G}'$ and Type-III $\hat{F}$, $\hat{G}$ and
$\hat{H}$.

We need to identify the $\tilde{U}$s in Eq.~(\ref{normalized_Yukawa})
as the $U_{m,\ell^{(\prime)}}^{f(\prime)}$ in
Eq.~(\ref{IC-Type-I})-(\ref{IC-Type-III}).
For Type-I in Fig.~\ref{fig:SGG-new-p=3/2-T1}, the relation between the
mixing coefficients of $\hat{F}$ 
and $\hat{F}'$ is given from Eq.~(\ref{IC-Type-I}) by 
\begin{align}
&\tilde{U}_1^f=\tilde{U}_{1,1}^f=1,\ \ \
\tilde{U}_2^f=\tilde{U}_{2,2}^f=
-\frac{x_F'\langle\phi_{F0}\rangle}{z_F'\langle\psi_{F-3/2}\rangle}
\frac{D_{0,0}^{\alpha',\alpha',S_F}}{D_{0,2}^{\alpha',\alpha,S_F'}},\\
&\tilde{U}_1^f=
\tilde{U}_{0,1}^{f\prime}=
-\frac{x_F\langle\phi_{F0}\rangle}{z_F\langle\psi_{F-3/2}\rangle}
\frac{D_{0,0}^{\alpha,\alpha,S_F}}{D_{0,1}^{\alpha,\alpha',S_F'}},\ \ \
\tilde{U}_2^f=\tilde{U}_{1,2}^{f\prime}=
-\frac{x_F\langle\phi_{F0}\rangle}{z_F\langle\psi_{F-3/2}\rangle}
\frac{D_{1,1}^{\alpha,\alpha,S_F}}{D_{1,2}^{\alpha,\alpha',S_F'}},
\end{align}
where 
$\tilde{U}_{1,1}^f:=U_{1,1}^f/U_{0,0}^f$,
$\tilde{U}_{2,2}^f:=U_{2,2}^f/U_{0,0}^f$,
$\tilde{U}_{0,1}^{f\prime}:=U_{0,1}^{f\prime}/U_{2,0}^{f\prime}$, and 
$\tilde{U}_{1,2}^{f\prime}:=U_{1,2}^{f\prime}/U_{2,0}^{f\prime}$.
For Type-II in Fig.~\ref{fig:SGG-new-p=3/2-T2}, 
the relation between the mixing coefficients of $\hat{F}$
and $\hat{F}'$ is given from Eq.~(\ref{IC-Type-II}) by 
\begin{align}
&\tilde{U}_1^f=\tilde{U}_{1,1}^f=1,\ \ \
\tilde{U}_2^f=\tilde{U}_{2,2}^f=1,\\
&\tilde{U}_1^f=\tilde{U}_{1,1}^{f\prime}=
\frac{D_{1,1}^{\alpha,\alpha,S_F}}{D_{0,0}^{\alpha,\alpha,S_F}}
\frac{D_{0,0}^{\alpha,\alpha',S_F'}}{D_{1,1}^{\alpha,\alpha',S_F'}},\ \ \
\tilde{U}_2^f=\tilde{U}_{2,2}^{f\prime}=
\frac{D_{2,2}^{\alpha,\alpha,S_F}}{D_{0,0}^{\alpha,\alpha,S_F}}
\frac{D_{0,0}^{\alpha,\alpha',S_F'}}{D_{2,2}^{\alpha,\alpha',S_F'}},
\end{align}
where
$\tilde{U}_{1,1}^{f(\prime)}:=U_{1,1}^{f(\prime)}/U_{0,0}^{f(\prime)}$
and 
$\tilde{U}_{2,2}^{f(\prime)}:=U_{2,2}^{f(\prime)}/U_{0,0}^{f(\prime)}$.
For Type-III in Fig.~\ref{fig:SGG-new-p=3/2-T3}, 
the relation between the mixing coefficients of $\hat{F}$
and $\hat{F}'$ is given from Eq.~(\ref{IC-Type-III}) by 
\begin{align}
&\tilde{U}_1^f=
\tilde{U}_{1,1}^{f\prime}=
\frac{D_{q_\alpha-1/2,q_\alpha-1/2}^{\alpha,\alpha,S_F}}
{D_{q_\alpha-3/2,q_\alpha-3/2}^{\alpha,\alpha,S_F}}
\frac{D_{q_\alpha-3/2,0}^{\alpha,\alpha',S_F'}}
{D_{q_\alpha-1/2,1}^{\alpha,\alpha',S_F'}},\ \ \
\tilde{U}_2^f=
\tilde{U}_{2,2}^{f\prime}=
\frac{D_{q_\alpha+1/2,q_\alpha+1/2}^{\alpha,\alpha,S_F}}
{D_{q_\alpha-3/2,q_\alpha-3/2}^{\alpha,\alpha,S_F}}
\frac{D_{q_\alpha-3/2,0}^{\alpha,\alpha',S_F'}}
{D_{q_\alpha+1/2,2}^{\alpha,\alpha',S_F'}},
\end{align}
where
$\tilde{U}_{1,1}^{f\prime}:=U_{1,1}^{f\prime}/U_{0,0}^{f\prime}$ and
$\tilde{U}_{2,2}^{f\prime}:=U_{2,2}^{f\prime}/U_{0,0}^{f\prime}$.
We can also get the coefficients
$\tilde{U}_{n,\ell^{(\prime)}}^{g(\prime)}$ 
by replacing $F$, $F'$ $f$, $\alpha$ and $\alpha'$ by $G$, $G'$, $g$,
$\beta$ and $\beta'$.

For example, we calculate the eigenvalues of the Yukawa coupling
constants for Type-II $\hat{F}^{(\prime)}$, $\hat{G}^{(\prime)}$ and
$\hat{H}^{(\prime)}$. 
For $S_H=0,S_H'=1/2,q_\gamma=1/2$,
we obtain 
\begin{align}
b_n^{h(\prime)}=
\sqrt{\frac{\Gamma(2\gamma^{(\prime)})}{n!\Gamma(2\gamma^{(\prime)}+n)}}.
\end{align}
This leads to
\begin{align}
&\tilde{y}_0\simeq 1,\nonumber\\
&\tilde{y}_1\simeq -
\tilde{U}_{1}^f\tilde{U}_{1}^g\epsilon_H^2
\frac{\sqrt{\alpha\beta}}{2\gamma^{(\prime)2}(2\gamma^{(\prime)}+1)},
\nonumber\\
&\tilde{y}_2\simeq
\tilde{U}_{2}^f\tilde{U}_{2}^g\epsilon_H^4
\frac{\sqrt{\alpha\beta(2\alpha+1)(2\beta+1)}}
{4\gamma^{(\prime)2}(\gamma^{(\prime)}+1)^2
(2\gamma^{(\prime)}+1)(2\gamma^{(\prime)}+3)}.
\end{align}
For $\alpha=\beta=\gamma/2=1/2$,
$\tilde{y}_0\simeq 1$,
$\tilde{y}_1\simeq -\epsilon_H^2/12$, and
$\tilde{y}_2\simeq\epsilon_H^4/240$.
For $\alpha=\beta=\gamma/2=1$,
$\tilde{y}_0\simeq 1$,
$\tilde{y}_1\simeq -\epsilon_H^2/40$, and 
$\tilde{y}_2\simeq\epsilon_H^4/1680$.

We find that the overall coupling can be suppressed by
the mixing coefficients  $U_{0,0}^{f\prime}$, $U_{0,0}^{g\prime}$, and
$U_{0}^{h\prime}$ of the Type-II $\hat{F}'$, $\hat{G}'$, and $\hat{H}'$,
respectively. 
For instance, we obtain for the matter field $\hat{H}$
from Eq.~(\ref{IC-Type-II}), 
\begin{align}
U_{0}^{h\prime}
\simeq \sqrt{\epsilon_H}
\frac{\Gamma(2\gamma+S_H)}{\Gamma(2\gamma+S_H'+q_\gamma)}
\sqrt{\frac{\Gamma(2\gamma')}{\Gamma(2\gamma)}
\frac{(S_H'+q_\gamma)!(S_H'-q_\gamma)!}{(S_H!)^2}},
\end{align}
when $U_{0}^{h\prime}=U_{0,0}^{\prime}$, and we assume $x_H/z_H\simeq
x_H'/z_H'$. 
We can calculate some examples of one generation.
For $S_H=0$, $S_H'=1/2$, $\gamma=1$, 
$U_{0}^{h\prime}\simeq -\sqrt{\epsilon_H}/\sqrt{2}$.
For $S_H=0$, $S_H'=1/2$, $\gamma=2$, 
$U_{0}^{h\prime}\simeq -\sqrt{3\epsilon_H}/2$.
For $S_H=0$, $S_H'=1/2$, $\gamma=4$, 
$U_{0}^{h\prime}\simeq -\sqrt{2\epsilon_H}/4$.
The other mixing coefficients of the matter fields $\hat{F}$ and
$\hat{G}$ can be calculated as well.
This is true for Type-III $\hat{F}'$ and $\hat{G}'$, where the
suppression factor is derived from Eq.~(\ref{IC-Type-II}).

We consider the following Yukawa coupling constants of Type-III
$\hat{F}$ and $\hat{G}$ $(q_\alpha,q_\beta>3/2)$, and Type-II $\hat{H}$,  
\begin{align}
y^{mn}=y U_0^{h}
\epsilon_H^{m+n+q_\alpha+q_\beta-3}
C_{m+q_\alpha-3/2,n+q_\beta-3/2}^{\alpha,\beta,0}
b_{m+n+q_\alpha+q_\beta-3}^{h},
\end{align}
where $U_{m,m+q_\alpha-3/2}^f=U_{n,n+q_\beta-3/2}^g=1$ $(m,n=1,2,3)$.
The ratio of the 00 component of the overall
coupling constant of the Type-III to that of the Type-II $\hat{F}$,
$\hat{G}$, and $\hat{H}$ is 
$\epsilon_H^{q_\alpha+q_\beta-3}
C_{q_\alpha-3/2,q_\beta-3/2}^{\alpha,\beta,0}
b_{q_\alpha+q_\beta-3}^h/C_{0,0}^{\alpha,\beta,0}b_{0}^h$,
where $C$s are given in Eq.~(\ref{CGC-C}),
and $b$s are given in Eq.~(\ref{mixing-coefficient-half-1}).
We find that
$C_{q_\alpha-3/2,q_\beta-3/2}^{\alpha,\beta,0}/
C_{0,0}^{\alpha,\beta,0}<1$, 
$b_{q_\alpha+q_\beta-3}^h/b_{0}^h<1$,
and $\epsilon_H^{q_\alpha+q_\beta-3}<1$ assuming $\epsilon_H<1$.
The overall coupling constant of Type-III $\hat{F}$ or $\hat{G}$
$(q_\alpha,q_\beta>3/2)$ is suppressed.

We comment on an overall constant of the Yukawa couplings. In the MSSM,
the value depends on $\tan\beta$. 
For $\tan\beta=(10,38,50)$, the overall coupling constant of
the up-type quark, the down-type quark, and the charged lepton is
$y^t\simeq(0.48,0.49,0.51)$, $y^b\simeq(0.051,0.23,0.37)$, 
and $y^\tau\simeq(0.070,0.32,0.51)$, respectively, at the GUT-scale
given by Ref.~\cite{Ross:2007az}. When we seek the 'tHooft's naturalness
\cite{'tHooft:1980xb} that the value of a dimensionless coupling
constant is $O(1)$ without any reason, the overall coupling constant
should be suppressed by some reasons for at least not large $\tan\beta$.
From the above discussion, the overall Yukawa coupling constant of the
Type-II or Type-III is naturally suppressed.
At present, it is difficult to say which type is better because we do
not know the value of $\tan\beta$.

\section{Summary and Discussion}
\label{summary}

We examined the model that contains structure fields with an integer
spin and a half-integer spin of the horizontal 
symmetry $SU(1,1)$. The model must include two sets of matter fields,
e.g., $\hat{F}(\hat{F}^c)$ and $\hat{F}'(\hat{F}^{c\prime})$, where
the weight value of the matter field $\hat{F}$ must be different by
a half-integer compared to that of the $\hat{F}'$. Although this model is
more complicated than the model that only includes two structure fields
with integer spins, we found that there are solvable vacuum
conditions that exactly determine the mixing coefficients of the chiral
modes $\hat{f}_n$ included in the matter fields $\hat{F}$ and
$\hat{F}'$. We classified the mixing coefficients of matter
fields into three types based on the component of the VEV of the
structure field and the difference between the values of the weights of
the matter fields. In addition, we determined that the model containing
the structure field with a half-integer spin can only generate odd
numbers of generations of matter fields at low energy as long as the VEV 
dominantly contributes to produce these generations.

We made sure that each Yukawa coupling constant of the chiral
particles at low energy is completely determined by the mixing
coefficients of the matter fields, the CGC of the matter fields and an
overall coupling constant. 
We examined a typical example of a Yukawa coupling matrix by using the
mixing coefficients obtained in \S \ref{sec:SGG}. 
We showed that the pattern of the Yukawa coupling constants 
in the model including the structure field with a half-integer spin
is not only hierarchical, but also can be different from that 
in the model only including the structure fields with integer spins,
because of the mixing between the matter fields, e.g. $\hat{F}$ and
$\hat{F}'$.
Smaller values for the $SU(1,1)$ spins of the structure fields are
preferred to produce the moderate hierarchy of quarks and leptons. 
Smaller values for the $SU(1,1)$ weights of the matter fields are also
preferred except in the marginal assignment.
In some cases, we pointed out that the overall coupling constant of the
Yukawa coupling is naturally suppressed. 

We also determined that the VEV of the $3/2$ component of a structure
field with a half-integer spin $S(\geq 3/2)$ can realize three
generations. The previous works in Refs.~\cite{Yamatsu:2007,Yamatsu:2008}
must contain the VEV of the $3$
component of a structure field with an integer spin $S(\geq 3)$ to realize
three generations. That is, the minimum value of $SU(1,1)$ spin is $S=3/2$
and $S=3$, respectively. 
We have not yet solved the vacuum structure in models with the $SU(1,1)$
spins $3/2$ or $3$ and we do not know which one is the more stable
choice, but we expect the field with spin $3/2$ to be much easier than
that with spin $3$ when we analyze these vacuum structures. 

We should mention the current status of noncompact gauge theory. 
One may wonder how to construct acceptable noncompact gauge theories
because such theories violate unitarity and have no stable vacuum in the
canonical forms, a K\"ahler potential $K=\hat{\Psi}^\dag\eta\hat{\Psi}$
and a gauge kinetic function $f_{AB}(\hat{\Psi})=\eta_{AB}$, which
include at least one indefinite metric.
The author has previously discussed an $\mathcal{N}=1$ supersymmetric
noncompact $SU(1,1)$ 
gauge theory \cite{Yamatsu:2009}. That the Lagrangian has all positive
definite metrics for the physical fields at the proper vacuum is
realized by  carefully selecting the K\"ahler potential, the gauge
kinetic function, and the superpotential. Unfortunately, this model only 
includes a chiral superfield with the $SU(1,1)$ spin 1. Thus, to obtain
the ``full''  Lagrangian, we must construct acceptable models
that produce three generations of quarks and leptons and one generation
of higgses.

\section*{Acknowledgments}

The author would like to thank Kenzo Inoue for useful suggestions and
stimulating encouragement, and Micheal S. Berger, Jonathan P. Hall,
Kentaro Kojima and Hirofumi Kubo for helpful comments. 
This work is supported in part by DOE grant DE-FG02-91ER40661.

\bibliographystyle{utphys} 
\bibliography{reference}

\end{document}